\documentclass[fleqn,10pt]{wlscirep}
\usepackage[utf8]{inputenc}
\usepackage[T1]{fontenc}
\usepackage{lineno}
\usepackage{rotating}
\title{A dataset of hourly sea surface temperature from drifting buoys}

\author[1,*]{Shane Elipot}
\author[2]{Adam Sykulski}
\author[3]{Rick Lumpkin}
\author[4]{Luca Centurioni}
\author[3]{Mayra Pazos}
\affil[1]{Rosenstiel School of Marine and Atmospheric Science, University of Miami, Miami, FL 33149, USA}
\affil[2]{Lancaster University, Department of Mathematics and Statistics, Lancaster, LA1 4YW, UK}
\affil[3]{NOAA Atlantic Oceanographic and Meteorological Laboratory, Miami, FL 33149, USA}
\affil[4]{Lagrangian Drifter Laboratory, Scripps Institution of Oceanography, University of California San Diego, San Diego, CA 92103, USA}

\affil[*]{corresponding author(s): Shane Elipot (selipot@miami.edu)}

\begin{abstract}
A dataset of sea surface temperature (SST) estimates is generated from the temperature observations of surface drifting buoys of NOAA's Global Drifter Program. Estimates of SST at regular hourly time steps along drifter trajectories are obtained by fitting to observations a mathematical model representing simultaneously SST diurnal variability with three harmonics of the daily frequency, and SST low-frequency variability with a first degree polynomial. Subsequent estimates of non-diurnal SST, diurnal SST anomalies, and total SST as their sum, are provided with their respective standard uncertainties. This Lagrangian SST dataset has been developed to match the existing and on-going hourly dataset of position and velocity from the Global Drifter Program.
\end{abstract}
\begin{document}

\flushbottom
\maketitle

\thispagestyle{empty}

%\noindent Please note: Abbreviations should be introduced at the first mention in the main text – no abbreviations lists or tables should be included. Structure of the main text is provided below.

\section*{Background \& Summary}

%\emph{(700 words maximum) An overview of the study design, the assay(s) performed, and the created data, including any background information needed to put this study in the context of previous work and the literature. The section should also briefly outline the broader goals that motivated the creation of this dataset and the potential reuse value. We also encourage authors to include a figure that provides a schematic overview of the study and assay(s) design. The Background \& Summary should not include subheadings. This section and the other main body sections of the manuscript should include citations to the literature as needed.}
%The GDP is part of NOAA's Global Ocean Observing System and is a scientific project of the international Data Buoy Cooperation Panel (DBCP, \url{http://www.jcommops.org/dbcp/}).

The Global Drifter Program (GDP) funded by the U.S. National Oceanic and Atmospheric Administration (NOAA) maintains an array of satellite-tracked water-following drifting buoys, hereafter referred to as \emph{drifters}, designed to acquire in situ observations of near-surface ocean current, sea surface temperature (SST), and atmospheric sea level pressure \cite{centurioni2019global}. The requirement of the Global Ocean Observing System (GOOS) to achieve a nominal $5^\circ \times 5^\circ$ coverage of the world's ocean has been fulfilled since September 2005 with a pool of 1250 drifters \cite{lumpkin2016fulfilling}. In near-real time, drifter locations and sensor data are relayed to the WMO's Global Telecommunication System (GTS), contributing to the collection of critical information needed for the World Weather Watch programme. Drifter data are also harvested by various national and international projects and organizations which aim at assembling in situ SST observations to produce quality-controlled and reformatted datasets for scientific {analysis}, climate monitoring, and calibration and validation of satellite-based SST observations. In delayed-time, the GDP maintains the historical database of drifter data and metadata, and delivers regular updates of drifter data products of surface currents and SST following quality control and estimation procedures. 
The historical observations, with the earliest ones from 1979, have been processed in incremental steps to generate a 6-hour joint dataset of drifter position, velocity, and SST estimates, along with their uncertainty estimates \cite{hansen1996quality,lumpkin2019noaa}. Because the frequency of drifter observations has increased since the onset of the array, an hourly product of drifter velocity estimates with uncertainties has been generated following a new estimation methodology, since 2016 \cite{elipot2016global}. 
This paper describes the methods that have now been devised to generate a new dataset of SST estimates at hourly time steps along drifters' trajectories, aimed at accompanying the on-going hourly drifter velocity dataset \cite{elipot2016global}. {The dataset of drifter position and velocity estimates augmented with SST estimates is now the version 2.00 of the Hourly location, current velocity, and temperature collected from Global Drifter Program drifters world-wide dataset\cite{elipot2022hourly}. A summary of the products generated by the GDP is contained in Table~\ref{tab:datamatrix}.

Hourly estimates of SST along drifters' trajectories are ultimately obtained from in situ sea water temperature observations. Estimates are obtained by least squares fitting a mathematical model of SST temporal evolution to temporally-uneven SST observations. The adopted fitting method is an adaptation of the locally weighted scatterplot smoothing method, known as LOWESS \cite{Cleveland1979}. The method operates in an iterative manner in order to gradually reduce originally-uniform weights given to observations, eventually rejecting observations diagnosed as outliers. The method first generates SST estimates at the original times of the drifter SST sensor observations, and second generates SST estimates at regular top-of-hour times that typically do not coincide with the observation times. After fitting the mathematical model, the local error variance of the assumed observational process is estimated by summing the variance of the residuals from the fit and an ad hoc term aimed at taking into account the quantization error arising from temperature sensor resolution \cite{chiorboli2003uncertainty}.  The ultimately chosen mathematical SST model is the sum of a polynomial function of order one, meant to capture non-diurnal variability, and the sum of three pairs of cosine and sine functions at harmonic frequencies of the diurnal frequency, meant to capture diurnal variability. The error variance estimates are subsequently propagated through the least squares method to derive standard uncertainties of the model parameter and of the SST estimates. The parameters of the chosen model and of the fitting method have been chosen by analyzing two limited subsets of the global observational dataset. The choices made aim at minimizing both the mean square error calculated from the residuals of the fits and the estimates of the error variance of the observational process model. Ultimately, the variables added to the existing hourly dataset of drifter positions and velocities consist of SST diurnal hourly estimates, SST non-diurnal hourly estimates, and total SST hourly estimates. Each of these estimates is accompanied by its respective standard error estimates and specifically devised quality flag. Global statistics from our estimates indicate that the square root of the typical error variance is 0.02$^\circ$C for drogued drifters' observations and 0.03$^\circ$C for undrogued drifters' observations, and that the typical standard error for SST estimates is 0.016$^\circ$C for drogued drifters' observations and 0.019$^\circ$C for undrogued drifters' observations. There exist, however, marked geographical differences for these values across the world's ocean. The magnitudes of these uncertainties are an order of magnitude smaller than previously estimated measurement uncertainties for drifting buoys \cite{kennedy2014review} because of differences of methods. % 707-11

\section*{Methods}

%\emph{The Methods should include detailed text describing any steps or procedures used in producing the data, including full descriptions of the experimental design, data acquisition assays, and any computational processing (e.g. normalization, image feature extraction). See the detailed section in our submission guidelines for advice on writing a transparent and reproducible methods section. Related methods should be grouped under corresponding subheadings where possible, and methods should be described in enough detail to allow other researchers to interpret and repeat, if required, the full study. Specific data outputs should be explicitly referenced via data citation (see Data Records and Citing Data, below).}

%Authors should cite previous descriptions of the methods under use, but ideally the method descriptions should be complete enough for others to understand and reproduce the methods and processing steps without referring to associated publications. There is no limit to the length of the Methods section. Subheadings should not be numbered.

The methods described in this paper define four levels of SST data denoted {Level-0,1,2,3}, as explained in the following sections:
\begin{itemize}
\item {Level-0 corresponds to the original, temporally unevenly distributed data as reported by the SST sensor and transmitted to the GDP DAC via Service Argos system or by the Lagrangian Drifter Laboratory at Scripps Institution of Oceanography;}
\item {Level-1 data result from the application of initial processing and quality-controls to Level-0 data;}
\item {Level-2 corresponds to SST estimates at the same unevenly distributed times as Level-1;}
\item {Level-3 corresponds to SST estimates at a regular hourly interval, at the top of each hour. Level-3 estimates are obtained at the same times as the position and velocity estimates for drifters of the of the GDP hourly dataset, release 1.04c} \cite{elipot2016global} {. The dataset of drifter position and velocity estimates augmented with SST estimates is release 2.00.}
\end{itemize}

{This paper describes the derivations of Level-1,2,3 datasets and announces the release of the Level-3 data as a product.}

\subsection*{Data collation}

In its basic configuration, a standard SVP drifter (from the Surface Velocity Program of the World Ocean Circulation Experiment) is composed of a surface float tethered to a holey-sock ``drogue'', or sea anchor, centered at 15~m depth when deployed. As a result, the surface displacement of the float tracked by satellites is predominantly representative of oceanic velocity at 15~m \cite{lumpkin2017advances}. With time, a drifter can lose its drogue and becomes ``undrogued''\cite{lumpkin2016fulfilling} but still continues to transmit its position and its sensor data until it dies \cite{lumpkin2012evaluating}. In addition to the standard SVP configuration, a number of drifters are equipped with additional sensors such as a barometer for sea-level atmospheric pressure\cite{centurioni2017global}  or a conductivity sensor to measure salinity\cite{centurioni2015sea} (see \url{https://www.aoml.noaa.gov/phod/dac/deployed.html} for the historical deployment log of the GDP). Yet, all drifters are equipped with a temperature sensor attached to the surface float and located at about 18~cm depth when at rest. Despite being an environmental variable of climate importance, sea surface temperature does not have a unique definition, and the depth at which a measurement is taken is crucial to interpret its value and variability. The definitions for near-surface seawater temperature from the Group for High Resolution Sea Surface Temperature (GHRSST, \url{https://www.ghrsst.org/ghrsst-data-services/products/}) would suggest to call the temperature data from surface drifters ``observations of sea water temperature at a depth of 18 cm''. In the rest of this paper, for simplicity, we will refer to temperature observations from drifters, as well as temperature estimates derived from these, as SST data. 

At the onset of the GDP in 1979, drifters were tracked by Argos which is both a positioning system and a data transmission system. At the beginning of the program, battery power and money {were} conserved by sampling location and transmitting data using different 1/3 and 2/3 schemes. As an example, data was transmitted for one day, followed by two days of no transmission, or data was transmitted for 8 hours, followed by 16 hours of no transmission \cite{hansen1989temporal}. Since 2000, this sampling scheme has been abandoned thanks to increased battery {life} and other technological advancements. At the same time, the number of operational satellites of the Argos constellation increased with time, so that the typical time interval between two consecutive Argos fixes reduced to between 1 and 2 hours \cite{elipot2008spectral}. However, stemming from the original sampling pattern, the GDP has continued to routinely process and interpolate the location fixes and temperature observations to produce drifter locations and temperature estimates continuously along trajectories at 6-hour intervals. The general method of interpolation, called kriging \cite{hansen1996quality}, provides an estimate of location, or of SST, at a given time as a weighted linear combination of observations close in time (the five previous ones and the five subsequent ones in this case). Finding the optimal set of weights involves assuming a mathematical expression for a so-called structure function which is half of the variance of observation differences as a function of temporal lag. For the 6-hourly GDP product, structure functions for location or temperature are fitted regionally and in discrete time periods to observations. As such, the kriging implementations for either location or temperature differ because of the structure function employed, and estimates of location and SST are independent from each other in the sense that no location information is used to estimate SST and vice versa. Drifter velocities are subsequently computed from the 6-hour locations by 12-hour central differencing \cite{hansen1996quality}. The GDP started to phase out the Argos positioning system in 2014 in favor of the Global Positioning System (GPS). This system provides locations with estimated $O(10)$-meter scale accuracy\cite{elipot2016global} that {are} relayed almost instantly via the Iridium Satellite Communication system, along with sensor data, at regular temporal interval (typically hourly but not always), in contrast to Argos locations and transmissions. At the time of writing, the transition to GPS tracking and Iridium transmission is complete. A few drifters of the array were equipped with GPS receivers that transmitted their data via the Argos system \cite{elipot2016global}.

All transmitted locations and sensor data from Argos-tracked drifters were collected by {Collecte} Localisation Satellite (CLS) which relayed them, first in near-real time to the WMO's GTS, and second to the GDP Data Assembly Center (DAC) located at the NOAA Atlantic Oceanographic and Meteorological Laboratory (AOML) in Miami, Florida. The Argos data received by the DAC are organized in messages, each associated with an Argos localization from a single Argos satellite pass, with a time stamp contained within the 10 to 20 min duration of that pass \cite{Argos,lopez2014improving}. Each message may contain one or more sets of sensor observations, each set having its own sensor time which differs from the localization time, but is typically within plus or minus the pass' duration. Sometimes, an Argos message does not contain a location and a location time but nevertheless contains some sets of observations. In this case, the DAC assigns the location and time of the previous message to these observations. The sensor observations are subsequently processed by the DAC as follows. In the case of a message containing multiple and distinct sets of observations, the median value of all observation times and SST observations are retained for that message. For some drifters with specific sampling configurations, observations may explicitly include an age, which is a time interval that needs to be subtracted from the nominal observation time to obtain the true time of observations.
Next, the DAC reorganizes Argos data as a row file, one per drifter, with each row containing in its columns Argos location (latitude and longitude), Argos location time, observation time, and observation data (SST and other sensors). Observation data are originally in sensor count but are decoded and converted throughout this process to physical units according to sensor equations found within each drifter specification sheet. Note that because several Argos satellites can be within the view of a single drifter at the same time, it is possible for the same set of observations to be transmitted by a drifter to different satellites, and to be eventually repeated in the dataset collated by the DAC, but with different locations and location times. As a result of the disconnection between Argos localization and acquisition of observations, there is no strict temporal coordination of location data and sensor data for Argos-tracked drifters. 

For the modern GDP drifters that relay their data through the Iridium Satellite System, the geographical location from a GPS receiver is treated as another sensor variable, like the sensor SST, and as such is not subject to the semi-aleatory transmission schedule like with the Argos system. The data are transmitted in Short Burst Data (SBD) format which contains a number of parameters that depends on the type of drifter and the manufacturer, but typically includes date and time and sensor data including GPS when available. GPS location times and sensor data times are therefore concurrent. If a GPS position is not available at an observation time, the previous position is reported with a recorded time delay. Depending on a drifter's firmware, the GPS location sampling interval may differ from the sensor data sampling intervals. Drifter locations and sensor data are relayed in near-real time to the GDP Data Processing Center (DPC) located at the Lagrangian Drifter Laboratory (LDL) of the Scripps Institution of Oceanography and, from there, are sent the  WMO’s GTS after decoding the GPS and sensor data according to manufacturers' specification sheets. The drifter messages decoded by the LDL are also made available as text files to the DAC at AOML for inclusion in the GDP database. {All data relayed by CLS and the GDP DPC, collated by the GDP DAC, form what we call here Level-0 data.}

%The Lagrangian Data Laboratory at Scripps Institution of Oceanography receives the SBD transmissions for the Iridium drifters, decodes the GPS and sensor data according to manufacturers' specification sheets, transmit these data in near real-time to the GTS, and makes them available as text files for the DAC to add to the GDP database. 

\subsection*{Pre-processing and initial quality controls}

%In fact, the method used here to derive SST estimates completely ignores the position information and uses only time as a coordinate. Only because the ultimate dataset of SST estimates are obtained at regular top-of-the-hour times, can they be associated to hourly position and velocity estimates. 

Out of the {Level-0 data}, we consider for the initial release of this new augmented product the SST observations from 20-Dec-1978 02:00:00 to 06-Jul-2020 22:59:31, which totals 285,886,818 data tuples of SST values and observation times. We apply a number of pre-processing and quality control procedures to these data to form {the Level-1} data. These initial procedures, as well as all subsequent estimation methods, are applied to all drifters irrespective of their tracking and data transmission systems. As we saw for Argos drifters, the nominal sampling patterns for drifter SST and location acquisition are essentially independent. Note that at this stage, as described in the previous section, an approximate or ``raw'' geographical location with varied, or unknown, uncertainty is associated with a SST data point.

The GDP DPC and DAC harvest drifter deployment sheets filled out at sea by operators, as well as conduct a number of diagnostics based on location and sensor data, in order to maintain a directory file at the DAC. This directory file lists the dates, times, and locations of trajectory starts, the dates, times, and locations of trajectory ends (i.e. drifter deaths), and the estimated dates and times of drogue losses \cite{lumpkin2012evaluating,lumpkin2013removing}. The start and death dates and times are used to truncate if needed the SST time series for data points before oceanic deployment and after oceanic death (``post-death'' sensor data may exist in the transmitted data for example if a drifter had been picked up by a vessel or run aground but continued to transmit its sensor data). Next, we {further determine blocks of time for which SST observations are deemed valid by applying} a quality-control step that has been in place as part of the production of the 6-hourly SST dataset. The NOAA Optimum Interpolation (OI) SST V2 \cite{huang2021improvments} at monthly time steps is used to calculate a monthly climatology which is subsequently interpolated to the raw locations associated with the drifter SST observations. These interpolated values are then visually compared to drifter SST observations to determine a first and a last ``good'' SST observation per drifter trajectory, based on an expert human assessment. The corresponding dates and times of these two points are recorded in a dedicated master file for all drifters. Another master file is maintained by the DAC that lists periods of time for which it appears that an SST sensor failed temporarily, also based on the comparison to climatological values. These potential periods of sensor failure, and the periods before and after the first and last good points, are subsequently discarded from the SST observation time series. Note that this comparison to a climatology is not used to remove seemingly outlying {individual} points, but rather to determine blocks of {invalid SST observations}. 
%We use the `tmpfl30.dat` and `badstmpfl.dat` files to determine beginning and ends of SST time series and periods of bad data to trim the raw data for level 0.
%Down to 245,172,611 data points.
After this stage, we remove some data records contain filling values for missing data points. 
%Down to 209,436,116 data points.
%This may have occurred because 
%Down to 197,917,558 data points.
Next, we find two or more SST observations existing at the same time for a single drifter. In that case, all observations are kept and will be processed for obtaining Level-1 estimates at the same times. 
%Note that there are SST observations at same times, 136,089 of them.
Next, we identify a number of drifter SST time series with only a single point, which are removed, and constant-value time series which are also removed. In the end, the final {Level-1} dataset consists of 197,916,695 tuples of SST and time data originating from 24,597 drifter trajectories.

As a result of the differing technologies of the data transmission systems (Argos and Iridium), of the number of different drifter manufacturers for the GDP, and of changing firmwares with time, the {Level-1} dataset of time series of SST observations is heterogeneous in its sampling intervals and apparent levels of noisiness. Two-dimensional histograms of occurrences of time differences and absolute temperature differences between two subsequent data points for all trajectories (Figure~\ref{fig:deltasst_deltat}) generally indicate that larger absolute SST differences are found for smaller time differences, for both Iridium and Argos drifters. For Iridium drifters, time differences are concentrated around multiples of one hour or 30 min, but with deviations from these because of possible delays of GPS signal acquisitions compared to a specified schedule. The distribution of time differences for Argos drifters is more continuous but exhibits local peaks near one hour and 101 min, the latter corresponding to the orbital period for an Argos satellite \cite{elipot2016global}. Even when considering the distribution of the median of time differences (or sampling interval) per trajectory, the Argos drifters exhibit a much {more} varied set of values compared to the Iridium drifters (Figure \ref{fig:deltat}).

\subsection*{Model of SST temporal evolution}

We seek to obtain SST estimates by fitting a local temporal model to the temperature data acquired along drifter trajectories. 
%First, the model is fitted to level-0 data to ultimately obtain level-1 SST estimates at the original observation times of level-0 data. Second, utilizing information gathered during the level-1 estimation, the model is fitted at regular hourly top-of-the-hour times to obtain level-2 SST estimates.
In-situ observations of SST from drifting or moored platforms, as well as remote sensing observations, suggest that two types of temporal variability typically co-exist: a relatively fast evolution on a diurnal time scale, sometimes referred to as a diurnal warming, and a relatively slower background evolution. For this background evolution (also referred to as non-diurnal in the rest of this manuscript), there is a priori no expectation of a dominant physical process acting at all times and all places. As such, it is reasonable to model this evolution with a local polynomial model as an approximation of a Taylor series expansion of an unknown underlying function \cite{fan1996local}. For the diurnal evolution, we follow a number of previous studies \cite{gentemann2003diurnal,kennedy2007global,lindfors2011climatological,morakbozzo2016climatological} and model this  evolution as the sum of cosine functions with fundamental frequency $\omega = 2\pi$ radians per day. In contrast to some previous studies however, our diurnal model is exactly periodic in the sense that it is locally zero-mean. A mean SST value and a possible difference of SST between the beginning and the end of a diurnal period will be both captured by the background non-diurnal polynomial model (as it will be at least of order 1). In addition, the amplitudes and phases of each of the cosine functions contributing to the diurnal model are not constant within a day, but rather vary locally since they are fitted at every time step using data within a sliding window centered on that time step. This diurnal model allows us to accommodate various environmental conditions (e.g. momentum and heat fluxes) affecting the shape of the diurnal signal in time and space as a drifter is advected by ocean currents. Note that since the diurnal SST estimate is locally zero-mean and does not represent solely a diurnal warming, the contemporaneous non-diurnal SST estimate differs from what is called a foundation temperature, that is a temperature free of diurnal temperature variability. In other words, the non-diurnal SST estimate typically contains the local mean of the SST diurnal variability.

In summary, the complete SST model is the sum of a polynomial $s_P$ of order $P$, and a sum $s_D$ of $N$ cosine functions at harmonic frequencies of the diurnal frequency $\omega = 2\pi$ radians per day. Next, we consider that a number of drifter SST observations $s_i$, found in the temporal vicinity of a single observation $s_k$ at time $t_k$, are generated by the process {described by the equation}
\begin{linenomath*}\begin{equation}
  \label{eq:model0}
  s_i = s_m(t_i;t_k) + \sigma_i \varepsilon_i,
\end{equation}\end{linenomath*}
{where $s_m(t_i;t_k)$ is the SST model and} the noise {model}, $\varepsilon_i$, is zero-mean, {has} unit variance [$E(\varepsilon_i) = 0$, $\text{Var}(\varepsilon_i) = 1$], and {is} independent of {the} noise at other times. The noise is locally scaled by the square root of $\sigma_i^2$ which is the error variance of the observations, is conditional to time $t_i$, and will be estimated a posteriori. 

The temporal evolution model is
\begin{linenomath*}\begin{align}                
                     s_m(t_i;t_k) & =  s_P(t_i;t_k) + s_D(t_i;t_k) \label{eq:model1}\\
                     & =   \sum_{p=0}^P s_{p,k} (t_i-t_k)^p + \sum_{n=1}^N A_{n,k} \cos[n \omega (t_i-t_k) + \phi_{n,k}] \label{eq:model2}\\
                     & =   \sum_{p=0}^P s_{p,k} (t_i-t_k)^p + \sum_{n=1}^N  \left[ \alpha_{n,k} \cos n \omega (t_i-t_k) +  \beta_{n,k} \sin n \omega (t_i-t_k) \right],\label{eq:model3}
\end{align}\end{linenomath*}
 with
\begin{linenomath*}\begin{eqnarray}
\label{eq:aphi}
\alpha_{n,k} & = & A_{n,k} \cos \phi_{n,k},\\
\beta_{n,k} & = & - A_{n,k} \sin \phi_{n,k}.
\end{eqnarray}\end{linenomath*}
The last form (\ref{eq:model3}) of the model shows that the $P+1+2N$ parameters of this model can be estimated by forming a linear system of equations. Ultimately, once the model parameters are estimated, the SST estimate itself at time $t_k$ is evaluated by setting $t=t_k$ in (\ref{eq:model3}) to obtain:
\begin{linenomath*}\begin{equation}
\label{eq:shat}
\widehat{s}_{m,k} \equiv s_m(t_k;t_k) =  s_{0,k} + \sum_{n=1}^N \alpha_{n,k}, %= s_{0,k} + \sum_{n=1}^N A_{n,k} \cos \phi_{n,k},
\end{equation}\end{linenomath*}
which involves only $N+1$ parameters of the $P+1+2N$ estimated parameters. The other $N+P$ parameters nevertheless provide further physical information such as the SST tendency for the non-diurnal evolution (e.g. $s_{1,k} = \partial s_P (t_k;t_k)/\partial t$ if $P \geq 1$) or the phase and amplitude of the diurnal harmonics:
\begin{linenomath*}\begin{eqnarray}
\phi_{n,k} & = & \arctan \left( \frac{-\beta_{n,k}}{\alpha_{n,k}}\right), \\
A_{n,k} & = & \frac{\alpha_{n,k}}{\cos \phi_{n,k}}.
\end{eqnarray}\end{linenomath*}

Ultimately, we will select $P=1$ and $N=3$ on the basis of the analyses of two subsets of surface drifters, as explained in the section \emph{Model selection}. As explained further in the next section, the model is first fitted at all original observation times 
%$t_k$, $k = 1,\ldots,K$, where $K$ is the total number of SST data points per 
of a drifter trajectory in a iterative manner in order to gradually adjust the weight of the data in the estimation, as well as identify outlier data points. After a given number of iterations, the model is ultimately fitted once at regular, top-of-the-hour, times that do not typically coincide with the original times.

\subsection*{Estimation of model parameters and SST}

The devised method to estimate SST continuously along a drifter trajectory is adapted from the method known as the locally weighted scatterplot smoothing or LOWESS \cite{Cleveland1979}. This method is iterative, and thus robust to outlying data points which are commonly observed in SST time series from surface drifters (see an example in Figure \ref{fig:example0}). Our method goes as follows. For a given SST time series from a single drifter, for each SST observation $s_k$ at time $t_k$, we compute by weighted least squares the $P+1+2N$ parameters of the model $s_m$ that minimize 
\begin{linenomath*}\begin{equation}
\label{eq:lsq}
\sum_{i=1}^K\left[ s_i - s_m(t_i;t_k) \right]^2 K_{h_k,i},
\end{equation}\end{linenomath*}
where $K_{h_k,i},$ is a set of weights given by
\begin{linenomath*}\begin{equation}
\label{eq:w}
  K_{h_k,i} = K\left( \frac{t_i - t_k}{h_k} \right),
\end{equation}\end{linenomath*}
with $K$ the tricube kernel function \cite{Cleveland1979}:
\begin{linenomath*}\begin{equation}
K(\tau) = (1-|\tau|^3)^3 I_{[-1,1]}(\tau),\quad \text{with} \quad I_{[-1,1]}(\tau) = \left\{\begin{array}{lr} 1, &  |\tau|\leq 1 \\ 0, & |\tau|> 1. \end{array} \right.
\end{equation}\end{linenomath*}
In~(\ref{eq:w}), $h_k$ is called the bandwidth of the kernel $K$, that is the half-width of the temporal window around the observation time $t_k$ within which the weights $K_{h_k,i}$ are different from zero. The least squares calculation therefore involves practically only those data points with non-zero weights. Because the complete model $s_m$ includes a diurnal oscillation, we initially set $h_k = 1$ day for all points, but this value is automatically and gradually increased as needed in 1-hour steps in order to include more data points to ensure that the least squares system of equations is not undetermined, up to a maximum value of 2 days. If not enough data points are available within the temporal window of maximum length, then no SST estimate is obtained. For the data selected to match the GDP hourly dataset version 1.04c (released in February 2021, with data through June 2020), fewer than 0.4\% of the data points require a half-bandwidth longer than 1 day.

Using matrix notation for convenience, the minimization problem (\ref{eq:lsq}) can be written 
\begin{linenomath*}\begin{equation}
\label{eq:minprob}
  \min_{\boldsymbol{\beta}} (\mathbf{s} - \mathbf{X} \boldsymbol{\beta})^T \mathbf{W} (\mathbf{s} - \mathbf{X} \boldsymbol{\beta}),
\end{equation}\end{linenomath*}
with solution
\begin{linenomath*}\begin{equation}
  \widehat{\boldsymbol{\beta}} = (\mathbf{X}^T\mathbf{W}\mathbf{X})^{-1}\mathbf{X}^T\mathbf{W}\mathbf{s}.
\end{equation}\end{linenomath*}
In (\ref{eq:minprob}), $\mathbf{X}$ is the design matrix for linear model~(\ref{eq:model3}):
\begin{linenomath*}\begin{equation}
\mathbf{X} = 
\begin{bmatrix} 
\mathbf{X}_1 & \mathbf{X}_2 & \mathbf{X}_3
\end{bmatrix},
\end{equation}\end{linenomath*}
with
\begin{linenomath*}\begin{equation}
\mathbf{X}_1 = 
\begin{bmatrix} 
1 & (t_1 - t_k) & \cdots  & (t_1-t_k)^P \\
\vdots & \vdots & & \vdots \\
1 & (t_i - t_k) & \cdots  & (t_i-t_k)^P \\
\vdots & \vdots & & \vdots \\
1 & (t_K - t_k) & \cdots  & (t_K-t_k)^P \\
\end{bmatrix},
\end{equation}\end{linenomath*}
\begin{linenomath*}\begin{equation}
\mathbf{X}_2 = 
\begin{bmatrix} 
\cos  \{\omega (t_1-t_k)\} & \cos  \{2\omega (t_1-t_k)\} &  \cdots & \cos \{N \omega (t_1-t_k)\} \\
\vdots & \vdots & &   \\
 \cos  \{\omega (t_i-t_k)\} &  \cos  \{2\omega (t_i-t_k)\} &  \cdots & \cos \{N \omega (t_i-t_k)\} \\
\vdots & \vdots & & \\
\cos  \{\omega (t_K-t_k)\} & \cos  \{2\omega (t_K-t_k)\} & \cdots & \cos \{N \omega (t_K-t_k)\} \\
\end{bmatrix},
\end{equation}\end{linenomath*}
\begin{linenomath*}\begin{equation}
\mathbf{X}_3 = 
\begin{bmatrix} 
\sin  \{\omega (t_1-t_k)\} & \sin  \{2\omega (t_1-t_k)\} &  \cdots & \sin \{N \omega (t_1-t_k)\} \\
\vdots & \vdots & &   \\
 \sin  \{\omega (t_i-t_k)\} &  \sin  \{2\omega (t_i-t_k)\} &  \cdots & \sin \{N \omega (t_i-t_k)\} \\
\vdots & \vdots & & \\
\sin  \{\omega (t_K-t_k)\} & \sin  \{2\omega (t_K-t_k)\} & \cdots & \sin \{N \omega (t_K-t_k)\} \\
\end{bmatrix}.
\end{equation}\end{linenomath*}
The {weight} matrix $\mathbf{W}$ is defined by
\begin{linenomath*}\begin{equation}
  \mathbf{W} = \text{diag}\left\{K_{h_k,i}\right\},
\end{equation}\end{linenomath*}
and $\mathbf{s}$ and $\boldsymbol{\beta}$ are the vector of data points and the vector of dimension $P+1+2N$ of parameters to be estimated, respectively:
\begin{linenomath*}\begin{equation}
\mathbf{s} = \begin{bmatrix}
 s_1 \\ \vdots \\ s_K  
\end{bmatrix},
\quad
\boldsymbol{\beta} = \begin{bmatrix}
 s_{0,k} \\ \vdots \\ s_{P,k} \\ \alpha_{1,k} \\ \vdots \\ \alpha_{N,k} \\ \beta_{1,k} \\ \vdots \\ \beta_{N,k}    
\end{bmatrix}.
\end{equation}\end{linenomath*}
 
Next, following the initial iteration of estimating the model parameters and calculating the corresponding SST estimates at all times $t_k$, we consider the residuals at all observation times:
\begin{linenomath*}\begin{equation}
  r_k = s_k - \widehat{s}_{m,k},
\end{equation}\end{linenomath*}
and compute $M$, the median value of the distribution of their absolute values. A set of robust weights are next calculated as
\begin{linenomath*}\begin{equation}
\label{eq:delta}
  \delta_k = B\left(\frac{r_k}{D M}\right),
\end{equation}\end{linenomath*}
where
\begin{linenomath*}\begin{equation}
  B(t) = (1-|t|^2)^2 I_{[-1,1]}(t),
\end{equation}\end{linenomath*}
is the biweight kernel function\cite{Cleveland1979} and $D$ is a real factor to be determined. 

The next step of the method consists in iterating the weighted least squares estimation of all parameters of the model at all times $t_k$, but this time using modified weights $\delta_i K_{h_k,i}$ instead of $K_{h_k,i}$ in~(\ref{eq:lsq}).  How many data points are down-weighted is dependent on the coefficient $D$ in the denominator of~(\ref{eq:delta}) which is typically set to 6 \cite{Cleveland1979} but here is set to 14, as discussed in the section \emph{Model selection}. The number of iterations is chosen here to be three after the initial least squares estimation without modified weights. The modified weights can effectively become zero when $\delta_k=0$, that is when the absolute value of a residual is larger than $D$ times $M$ for a given SST time series associated with one drifter trajectory. This implies that such data points are ultimately not used for any estimation but SST values at the corresponding time are nevertheless obtained using all available non-zero-weighted data points within the temporal window centered on any of these points. This method effectively flags as outliers some of the {Level-1} SST data point like a ``de-spiking'' procedure would do, for example by applying a median filter \cite{hansen1996quality}. An example of flagged outliers in a {Level-1} drifter SST time series is shown in Figure~\ref{fig:example0}. One implicit assumption of using~(\ref{eq:delta}) to modify the weights of the data is that all residuals from a given drifter SST time series originate from a common distribution, or equivalently that the statistics of the observations are constant within a given trajectory. This assumption may be violated if an entire drifter trajectory is long enough to experience environmental condition changes, or the characteristics of the SST sensor changes in an undetected fashion. An illustration of a {Level-2} estimation step is provided in Figure~\ref{fig:example1}.  

Finally, as the last step of the method, the SST model (\ref{eq:model3}) with the same number of parameters is fitted to the data but at times $t_k$ corresponding to the top of the hour UTC (00:00, 01:00, etc.), in one iteration with weights $\delta_i K_{h_k,i}$ where the $\delta_i$ were calculated prior to the last iteration for the original data times (not posterior). As before, the bandwidth is set to 1 day but is allowed to increase in increments of one hour, up to two days, to make sure the linear estimation problem is not under-determined. This last step generates the final {Level-3} data product. An illustration of {Level-3} estimated data is provided in Figure~\ref{fig:example2}.

\subsection*{Error variance estimates and uncertainty estimates}
% how we derive and interpret the uncertainties

As part of the method, we quantify the uncertainties of the parameter estimates and thus the uncertainties of the diurnal SST estimates, of the non-diurnal SST estimates, and of the total SST estimates.  Formally, the covariance matrix of the weighted least squares solution at time $t_k$ is 
\begin{linenomath*}\begin{equation}
\label{eq:varbeta}
%\text{Var}(\widehat{\boldsymbol{\beta}}) 
\mathbf{C}_{\boldsymbol{\beta}} \equiv  \text{Var}(\widehat{\boldsymbol{\beta}})  = (\mathbf{X}\mathbf{W^*}\mathbf{X})^{-1}(\mathbf{X}^T \mathbf{W^*} \boldsymbol{\Sigma} \mathbf{W^*} \mathbf{X})(\mathbf{X}\mathbf{W^*}\mathbf{X})^{-1},
\end{equation}\end{linenomath*}
where $\mathbf{W^*}$ is the weight matrix containing in its diagonal the modified weights $\delta_i K_{h_k}(t_i-t_k)$ from the penultimate iteration of the least squares estimation, and $\boldsymbol{\Sigma}$ is the unknown covariance matrix of the observation errors from the process model (\ref{eq:model0}). In order to proceed, we assume local homoscedasticity and that the errors are independent which results in $\boldsymbol{\Sigma} = \sigma^2(t_k) \mathbf{I}$, where the local error variance $\sigma^2(t_k)$ is unknown and needs to be estimated. In the case of a local polynomial regression of order $P$, it is recommended \cite{fan1996local} to re-conduct a polynomial fit of order $P+2$ and {to estimate} the error variance from the residuals of that fit. In our case, which is not a sole polynomial regression since model (\ref{eq:model1}) also includes trigonometric functions, the optimal course of action is unclear. Yet, to proceed, we classically calculate a first estimate of the error variance from the normalized weighted residual sum of squares:
\begin{linenomath*}\begin{align}
\label{eq:wrss1}
  \widehat{\sigma}^2_1(t_k) & = \frac{(\mathbf{s} - \mathbf{X} \widehat{\boldsymbol{\beta}})^T \mathbf{W^*} (\mathbf{s} - \mathbf{X} \widehat{\boldsymbol{\beta}})}{\text{tr}\left\{ \mathbf{W^*} - \mathbf{W^*}\mathbf{X}(\mathbf{X}^T\mathbf{W^*}\mathbf{X})^{-1}\mathbf{X}^T\mathbf{W^*}\right\}}\\
\label{eq:wrss2}
& = \frac{\Sigma_i \left[s_i - \widehat{s}_m(t_i;t_k)\right]^2 \delta_i K_{h_k}(t_i-t_k)}{\nu}.
\end{align}\end{linenomath*}
The denominator of (\ref{eq:wrss1}), referred to as $\nu$ in (\ref{eq:wrss2}), is the effective number of degrees of freedom for the residuals for weighted least squares \cite{fan1996local}. For ordinary least squares, $\nu$ would simply be the number of data points used to calculate $\widehat{\boldsymbol{\beta}}$ minus the number of parameters of the model ($P+1+2N$), but for weighted least squares cases, $\nu$ is smaller. 

The majority of drifters from the GDP database are equipped with temperature sensors returning a bit count $n$ used to calculate SST following the sensor equation: 
\begin{linenomath*}\begin{equation}
\label{eq:sstequation}
SST = a n + b, 
\end{equation}\end{linenomath*}
where $a$ is the resolution of the temperature sensor.  As such, the {Level-1} data should be rounded due to the resolution of the instrument recording. In the signal processing literature this is known as quantization, and has the effect of removing high resolution information in the data. As a result, the estimated error variance [$\widehat{\sigma}^2_1(t_k)$, (\ref{eq:wrss1})] should be increased to reflect the additional uncertainty created through quantization, as this information cannot be recovered. In the extreme case that the input data is the same value within a full window length then the increase to the error variance is $a^2/12$ \cite{chiorboli2003uncertainty}, following from the properties of the uniform distribution. As a result, adjusting for resolution, our total error variance is
\begin{linenomath*}\begin{equation}
\label{eq:wrss3}
  \widehat{\sigma}^2(t_k)  = \widehat{\sigma}^2_1(t_k)  + \frac{a^2}{12}. 
\end{equation}\end{linenomath*}
This adjustment is conservative, in that the effect of resolution {on} the error variance will decrease as the input values have more variance \cite{chiorboli2003uncertainty}. For simplicity, we use the conservative adjustment proposed above as this ensures the reported standard errors always include the resolution effect which should not be ignored. 

% This first estimate of the error variance however does not include the error induced by the quantization of the level-0 data, or in other words, the error due to the discrete resolution of the temperature sensors. The majority of drifters from the GDP database are equipped with temperature sensors returning a bit count $n$ used to calculate SST following the equation 
% \begin{linenomath*}\begin{equation}
% \label{eq:sstequation}
% SST = a n + b, 
% \end{equation}\end{linenomath*}
% where $a$ is the drifter-dependent resolution of the temperature sensor. The quantization error is here assumed to be an additive noise uniformly distributed in the interval $(-a/2,a/2)$ with zero mean and variance $a^2/12$ \cite{chiorboli2003uncertainty}. As such, the final error variance estimate is
% \begin{linenomath*}\begin{equation}
% \label{eq:wrss3}
%   \widehat{\sigma}^2(t_k)  = \widehat{\sigma}^2_1(t_k)  + \frac{a^2}{12}. 
% \end{equation}\end{linenomath*}

For about 85\% of the drifters, representing 83\% of the {Level-2} estimates, the resolution $a$ can be obtained from the individual specification sheets provided by the manufacturers. We identify in this way 179 different resolutions, ranging from 0.00260877$^\circ$C  to 0.17$^\circ$C. The three most common resolutions are 0.01$^\circ$C, 0.05$^\circ$C, and 0.08$^\circ$C. For the remaining 15\% of the drifters, some have an SST equation which is not a linear function of a sensor single bit count and the impact of the quantization error cannot be simply modeled as in (\ref{eq:wrss3}). Some other drifters have an unknown resolution because of the lack of available metadata. For these drifters, we estimate the resolution from the data as follows: we consider the time series of absolute SST {temporal} difference, bin these differences in 0.001$^\circ$C bins, and assign the resolution to the most common value that is not zero. In this way, the three most commonly estimated resolutions are 0.05$^\circ$C, 0.08$^\circ$C., and 0.043$^\circ$C. This method is successful in 92\% of cases when tested on the data of the drifters for which the resolution is known from the metadata. The overall distribution of all resolution values, as well as their temporal distribution, are illustrated in Figure~\ref{fig:resolution}.

We found it necessary to consider the resolution error for two reasons. First, since we have allowed our estimation algorithm to obtain a solution with as {few} data points as the number of model parameters to be estimated, and because of numerical precision errors, we find a small number of instances (0.33\% of the {Level-2} data) for which $\nu$, and therefore the first estimated error variance $\widehat{\sigma}^2_1(t_k)$, is small and negative. These instances are resolved by adding the resolution error.
%The corresponding SST estimates are consequently flagged for failed estimation of their uncertainties, as explained in section \ref{sec:quality-indication}.
Second, in some other instances (0.22\% of the {Level-2} results, see Figure~\ref{fig:deno_errorvar}), we find that the residuals, and hence the estimated error variance and the parameter uncertainties, are locally zero within numerical precision despite an ample number of data points available for the estimation. This occurs when the {Level-1} SST data does not change in value within the estimation window for reasons which are not clear. Once again, these instances are resolved by adding the resolution error, resulting in more realistic error estimates. Nevertheless, these two instances define two populations of the results that are clearly separated within a two-dimensional histogram of $\widehat{\sigma}^2_1(t_k)$ and $\nu$ (Figure~\ref{fig:deno_errorvar}). As such, we can flag these results using an empirical and ad-hoc condition: 
\begin{linenomath*}\begin{equation}
  \label{eq:flag}
  \log_{10} [\widehat{\sigma_1^2}]^{1/2} < -\frac{1}{2}\log_{10} |\nu| -10.
\end{equation}\end{linenomath*}

The final error variance estimate $\widehat{\sigma}^2(t_k)$ (\ref{eq:wrss3}) is a function of time $t_k$ and specific to a drifter because of the sensor resolution. This estimate is subsequently used to calculate an estimate of the local covariance matrix of the observations $\widehat{\boldsymbol{\Sigma}} = \widehat{\sigma}^2(t_k) \mathbf{I}$ and to calculate the covariance matrix $\mathbf{C}_{\boldsymbol{\beta}}$ [expression~(\ref{eq:varbeta})].

From the expression for the SST estimate (\ref{eq:shat}), its variance is
\begin{linenomath*}\begin{eqnarray}
  \label{eq:varhat1}
  \sigma^2_m \equiv \text{Var}\left[s_m(t_k;t_k)\right] & = & \text{Var}\left[ s_{0,k} + \Sigma_{n=1}^N \alpha_{n,k}\right]\\
  \label{eq:varhat2}
%& = & E\left[ \left(s_{0,k}\right)\left(s_{0,k}\right) \right] + 2 E\left[\left(s_{0,k}\right) \left(\Sigma_{n=1}^N \alpha_{n,k}\right)\right] + E\left[ \left(\Sigma_{n=1}^N \alpha_{n,k}\right) \left(\Sigma_{n=1}^N \alpha_{n,k}\right)\right]\\
& = & \text{Var}\left[ s_{0,k}\right] + 2 \text{Cov}\left[\left(s_{0,k}\right), \left(\Sigma_{n=1}^N \alpha_{n,k}\right)\right] + \text{Var}\left[ \Sigma_{n=1}^N \alpha_{n,k} \right]\\
  \label{eq:varhat3}
& = & \sigma^2_P + 2 \text{Cov}\left[\left(s_{0,k}\right) \left(\Sigma_{n=1}^N \alpha_{n,k}\right)\right] + \sigma_D^2.
\end{eqnarray}\end{linenomath*}
This last expression describes how the variance of the total SST estimate ($\sigma^2_m$) is the sum of the variance of the non-diurnal SST estimate ($\sigma^2_P$, containing one term), of the variance of the diurnal estimate ($\sigma^2_D$, containing $N^2$ terms), and of $2N$ additional cross-covariance terms. The $(N+1)^2$ needed terms to estimate $\sigma^2_m$, $\sigma^2_P$, and $\sigma^2_D$  are extracted and summed appropriately from the calculated covariance matrix $\mathbf{C}_{\boldsymbol{\beta}}$ [expression (\ref{eq:varbeta})] at each time step. The square root of each of these three estimated variances, referred {to} subsequently as $\widehat{\sigma}_m$, $\widehat{\sigma}_P$, and $\widehat{\sigma}_D$, define the standard errors, or standard uncertainties, of the three SST estimates. % [this to check again what I mean] As seen from expression (\ref{eq:varhat3}) because of cross-covariance terms, the standard error of the total SST estimate is always smaller (for 99.8\% of Level-1 data) or equal to the sum of the standard errors of the diurnal and non-diurnal SST estimates. 
Illustrations of estimated square roots of error variances and SST uncertainties is provided in Figures~\ref{fig:example1}, \ref{fig:example2}, and \ref{fig:example3}. {These figures, and Figure} \ref{fig:example3} {in particular, illustrate that the uncertainty estimates are temporally correlated for each individual drifter. Because the estimation procedure takes place within a sliding window, one should expect correlation at least among uncertainty estimates separated in time by less than the total length of the window (or twice the bandwidth} $h_k${, typically 2 days).} Additional discussions of error variance and uncertainty estimates are provided in section \emph{Interpretation of uncertainty estimates}.

%The appropriate subset of this matrix is summed for each time $t_k$ for all terms contributing to the diurnal SST standard error estimate $\widehat{\sigma}_D$ (1 term), to the non-diurnal SST standard error estimate $\widehat{\sigma}_P$  ($3 \times 3 = 9$ terms), and to the total SST standard error estimate $\widehat{\sigma}_m$ ($4 \times 4 = 16$ terms). 

\subsection*{Model selection}
\label{sec:model-selection}

In order to fit the total SST model to the data, choices need to be made for the order $P$ of the polynomial of the non-diurnal model and the number $N$ of harmonics of the diurnal model.  We consider a total of 14 models with $P$ varying between 0 and 3, and $N$ between 2 and 6 (Table \ref{tab:models}). We test and assess the performances of the models by fitting them to two limited subsets of GDP drifters as it would be computationally prohibitive to conduct tests on the entire {Level-1} data. 

The first subset is from the Salinity Processes in the Upper Ocean Regional Study (SPURS) in the subtropical North Atlantic \cite{centurioni2015sea,hormann2015evaluation}. The drifters released as part of SPURS were manufactured by Pacific Gyre Inc. but differed from standard SVP-type drifters of the GDP \cite{lumpkin2017advances}. Instead of a temperature sensor on their buoys, these drifters were equipped with an unpumped Sea-Bird Electronics SBE37-SI MicroCAT CTD placed underneath the surface buoy with its sensors located at a depth of 50 cm. The Microcat instruments were set to acquire conductivity and temperature at 30-min intervals by sampling once a minute for 5 min and averaging the values. According to the manufacturer, the initial accuracy and resolution of the temperature sensor are 0.002$^\circ$C and 0.0001$^\circ$C respectively, but the data transmitted and relayed to the DAC exhibit a resolution of 0.01$^\circ$C. For this study, we select 80 drifters which generated temperature data (considered to be SST observations) for time periods spanning between 29 and 660 days. These drifters transmitted their locations and sensor data via the Argos satellite system, including their position data from GPS receivers. These GPS data were previously used as a test set to devise the methodology being used to generate the global dataset of hourly position and velocity for the GDP \cite{elipot2016global}. However, here, the original Argos message data files for these drifters are re-processed to eliminate redundant and corrupted data by taking into consideration a previously ignored checksum flag indicating the integrity of Argos data transmissions. Next, the SST time series are further truncated to match the beginnings and ends of the  regular hourly time series of position and velocity for these drifters, as well as truncated for their first and last good data points as diagnosed by the QC procedures of the DAC. The resulting dataset consists of 80 time series of SST at uneven temporal intervals {(}multiple of 30 minutes{)}, totalling nearly 1.26M data points over 29,018 drifter days. 

%This dataset is however geographically limited to the tropical North Atlantic Ocean, and its temporal resolution is not fully representative of the entire dataset as xx\% of the drifter SST median sampling is near 1 hour (see Figure \ref{fig:}). 
The second subset of drifter SST data, hereafter referred to as the ``test'' subset, is built from the global database by selecting at random 14 drifters within each 10$^\circ$ latitude band between -70$^\circ$S and 70$^\circ$N with an average SST temporal sampling interval of between 50 and 70 minutes, resulting in a total of 98 individual SST time series which are further truncated in time for deployment times etc. The resulting dataset consists of 98 time series of SST at uneven temporal intervals, totalling 697,045 data points over 29,408 drifter days. These test drifters constitute a limited subset but represent a variety of drifter types deployed between years 2000 and 2019. Fifty of them are drifters with barometer (SVPB type), and 48 are standard SVP drifters. Fifty-seven of them were Iridium drifters and 41 Argos drifters. The test drifters were built by a variety of manufacturers: 18 by DBi, 19 by Metocean, 9 by Clearwater, 30 by Pacific Gyre, 18 by Scripps Institution of Oceanography, 2 by Technocean, 1 by Marlin-Yug, and 1 by NKE. Finally, the stated resolution of their SST sensors as specified by their respective specification sheets were varied: 0.01$^\circ$C for 68 of them, 0.05$^\circ$C for 20, 0.04$^\circ$C for 2, 0.043$^\circ$C for 2, 0.04329$^\circ$C for 1, 0.04343$^\circ$C for 1, 0.08$^\circ$C for one, and unknown for three of them.
%Yet, the numerical resolutions of the data collected at the GDP DAC for these drifters are 0.001 (4 of them), 0.005 (18), 0.01 (68), and 0.02 (2), 0.05 (5), and 0.1 (5).

We proceed to fit the 14 models listed in Table~\ref{tab:models} to the SPURS and test subsets of drifter SST time series, and subsequently consider two statistics calculated for each time series. The first statistic is the weighted root mean square error (WRMSE) calculated from the residuals of a given fit. For this calculation, the weights are the robust weights calculated by the algorithm described previously after the penultimate iteration (that is the weights calculated before the last estimation at the original times), but with a further normalization to ensure that weights sum to one:
\begin{linenomath*}\begin{align}
\label{eq:wrms}
  \text{WRMSE} & =  \left[ \sum_{k=1}^M w_k (s_k - \widehat{s}_{m,k})^2\right]^{1/2},\\
\label{eq:wj}
  w_k  & =  \frac{\delta_k K_{k,k}}{\sum_{i=1}^M \delta_i K_{i,i}}.
\end{align}\end{linenomath*}

%The second metric is the median of the standard errors for all estimates of a time series as an aggregate measure of both the original data temporal density and variance. 
The second statistic considered is the square root of the weighted median of the error variance estimates: after the final iteration, we consider the error variance estimates $\widehat{\sigma}^2(t_k)$ given by (\ref{eq:wrss3}) and using the weights defined by (\ref{eq:wj}), we calculate the weighted median defined as the value $\widehat{\sigma}^2(t_n)$ such that
\begin{linenomath*}\begin{equation}
\label{eq:wmedian}
\sum_{k=1}^{n-1} w_k \leq 1/2 \quad \text{and} \quad \sum_{k=n+1}^M w_k \leq 1/2.
\end{equation}\end{linenomath*}

For these two statistics, using weighted calculations effectively filters out the outlier data points diagnosed from the method (that for which $w_k=0$). We also tried non-weighted calculations that include all data points: besides shifting numerical values of the results in the sense of worsening performances, this did not change the relative performances of the models nor our overall conclusions and model selection choice.

%Considering these two statistics allow us to assess the performances of the model jointly in terms of bias and variance of the estimates. A model with more parameters could produce a closer fit to the data and reduce the bias and thus the WRMSE but is computationally more expensive, and may adversely increase locally the variance of the estimates which is dependent on the estimates of the error variance for which we compute the weighted median value. 
We find that varying the number of parameters of either the diurnal model or the non-diurnal model affects the two statistics differently. We present the results in Figure~\ref{fig:modelperfs}, displaying the WRMSE and the square root of the weighted median error variance, both in unit of degrees Celsius. The figures display scatter plots of the two statistics averaged over each of the subsets, along with ellipses representing the 95\% confidence intervals for the means in order to illustrate the scatter of the results. Not surprisingly, the scatter of the results is relatively smaller for the SPURS subset for which the SST records have the same nominal characteristics compared to the test subset composed of heterogeneous records.

For both data sets, we find that for a fixed number of harmonics of the diurnal model, increasing the polynomial order of the non-diurnal model (going right through the columns of Table \ref{tab:models}) reduces the error variance with little change to the WRMSE. The most dramatic reduction occurs when going from models for which $P=0$ (models 1, 2, and 3) to models for which $P=1$ (models 4 and higher) for which the square root of the weighted median error variance is at least approximately halved. Conversely, for a fixed order of the polynomial non-diurnal model, we find that increasing the number of harmonics (going down the rows of Table \ref{tab:models}) decreases the WRMSE with little change to the error variance. Considering these two general tendencies together, as well as the scatter of the results as depicted by the ellipses, we find that model 5 ($P=1$ and $N=3$) provides a good balance between the two statistics. Further, we find that from model 5, no significant improvement is obtained for the WMRSE error variance by increasing the polynomial order from 1 to 2 (going to model 8), and no significant improvement is obtained for the error variance by increasing the number of harmonics from 3 to 4 (going to model 6). Significant improvements are obtained for both statistics by both increasing the polynomial order from 1 to 2 and the number of harmonics from 3 to 4 (going from model 5 to model 9) for the SPURS subset but not for the test subset, which is expected to be representative of a much greater fraction of the total data. We also tested models with $P=1$ and $N=5, 6$ (models 13 and 14) but these, while reducing significantly the WRMSE from model 5, did not reduce significantly the error variance, and started to show larger error variances for the test subset. As a result, model 5 is our final choice of model to be fitted to the entire SST drifter dataset to generate the {Level-2} and {Level-3} datasets.

We now discuss briefly the choice of the bandwidth length [$h_k$, (\ref{eq:w})] and the choice of the factor $D$ for the robust weights [see~(\ref{eq:delta})]. The sensitivity of the results to these choices is summarized in Figure \ref{fig:modelperfs_test_2} for model 5 only. The choice of $h_k$ technically implies that data points within a $2 h_k$ window centered on the estimation time are considered [eqs.~(\ref{eq:lsq}) and (\ref{eq:w})]. Yet, because the weighing window is not uniform but a tricube kernel, the effective number of degrees of freedom used for each estimation is closer to the number of data points one would find in a uniform window of length $h_k$. Here, our choice $h_k=1$ day is based on observations that the characteristics of diurnal SST oscillations change on a daily time scale \cite{kawai2007diurnal}. Yet, we examine the summary statistics for model 5 for $h_k$ varying between 0.25 and 1.25 days at 0.25 day interval for the test subset (Figure \ref{fig:modelperfs_test_2}). We find that decreasing $h_k$ to less than 1 day significantly decreases the error variance yet does not decrease the WRMSE, and thus does not overall improve the performances of model 5. In contrast, we find that increasing $h_k$ to 1.25 days significantly increases the WRMSE and increases the error variance. The results for the SPURS subset are similar (not shown). These overall results therefore suggest that $h_k=1$ day is an appropriate choice for the bandwidth.

The choice of the factor $D$ in the denominator of the biweight kernel for calculating the robust weights [eq.~(\ref{eq:delta})] effectively sets the threshold for labeling data points as outliers. Our final choice of $D=14$ is compared to alternatively choosing $D$ between 4 and 20 at intervals of 2. The sensitivity of the summary statistics to the value of $D$ is displayed in Figure \ref{fig:modelperfs_test_2} for the test subset. We find that varying $D$ has a modest impact on the performances of model 5 and only for $D$ less than 6 does model 5 exhibits significantly better WRMSE, but no better standard deviation error. We also consider the ensemble average of the fraction of data points not labeled as outliers as a function of the choice of $D$ (Figure~\ref{fig:modelperfs_test_2}), and find that this fraction starts to decrease strongly as $D$ decreases from 8. In the original LOWESS method\cite{Cleveland1979}, $D$ is set to 6 without justification, and such a choice in our case would result in around 10\% of the data points labeled as outliers. In the end, we settled on $D=14$ which results in only between 1\% and 4\% of the data points being labeled as outliers, but maintains approximately the performance of model 5 compared to $D=6$. The results are similar for the SPURS subset (not shown).

\subsection*{Quality indication}
\label{sec:quality-indication}

The {Level-3} data product is intended to provide SST estimates contemporaneous to the estimated positions and velocities of drifters at hourly top-of-the-hour times from the hourly GDP dataset version 1.04c \cite{elipot2016global}, which with SST now included we shall call version {2.00}. Since the sampling of SST sensors onboard drifters can be independent from the positioning, it sometimes occurs that our methodology is able to provide an SST estimate at times when no location estimate is available. Since there is little use for an SST estimate with no associated location estimate, these are not included in the {Level-3} data product (see Table~\ref{tab:datamatrix}). 

We devise three different quality indication flag schemes, one for each component of SST (non-diurnal, diurnal, total), with flag values ranging from 0 (worst) to 5 (best), {with the intention of characterizing an increasing level of correctness}. For all three schemes, when no SST estimate could be obtained from the methodology (for example for lack of enough {Level-1} data within the sliding temporal window), or when SST data was simply not transmitted by a drifter (as an example because of a faulty sensor), the estimate is assigned quality flag 0 (and the NetCDF file contains a standard filling value). When an SST estimate could be obtained but not an SST uncertainty estimate, the estimate is assigned quality flag 1 (and the NetCDF file contains a filling value for the uncertainty estimate). 

For higher flag values, the schemes for the non-diurnal SST estimates and for the total SST estimates are the same, as illustrated in Figure~\ref{fig:quality_diagram}. When an SST estimate and an uncertainty estimate both exist, the quality flag is based on the relative position of the interval formed by the SST estimate plus or minus its standard error estimate with respect to the [-2,50]$^\circ$C range of physically-acceptable temperature values \cite{merchant2019satellite}. If the estimated interval is completely contained within this range, the assigned quality flag is the highest, at 5. If one or two end points of the interval are located outside of the range but the SST estimate is inside the range, the assigned quality flag is 4. If the SST estimate is outside the range but one of the end point of the interval is within the range, the assigned quality flag is 3. Finally, if the interval is located completely outside the physical range, then the quality flag is 2. {For analyses of total and non-diurnal SST, only estimates with quality flag 4 and 5 should be utilized. Estimates with quality flag 1, 2, and 3 are suspect and should not be used. They suggest that their corresponding true SST value are located outside of the physically-acceptable range. These estimates are nevertheless retained in the dataset with their distinct flags for traceability of the methods.}

The quality flag scheme for the diurnal SST estimates differs from the scheme described above because a diurnal SST estimate is an anomaly around zero for which a range of physically plausible values is not straightforward to define. A climatology of SST diurnal variability \cite{morakbozzo2016climatological} constructed by fitting a model to temperature observations from drifters within zonal bands, by seasons, and by environmental categories (clear or cloudy sky, wind speed) provides amplitude of SST diurnal anomalies no larger than 2.4$^\circ$C (from observations) or 0.689$^\circ$C (from modeled values). {Yet}, diurnal warming as large as 6.6$^\circ$C has been be detected {in coastal regions}\cite{flament1994amplitude}. {As a consequence}, rather than defining here an acceptable amplitude threshold for diurnal SST anomalies, we consider three criteria for the quality flag of a diurnal SST estimate: 
\begin{enumerate}
\item Is the {absolute value of the} diurnal anomaly estimate {strictly larger} than its standard error estimate? 
\item Is the standard error estimate for the diurnal estimate smaller than 1$^\circ$C?
\item Were more than 24 {Level-1} data points used to obtain an estimate? 
\end{enumerate}
As illustrated in Figure \ref{fig:sst2_esst2}, criteria (1) and (2) define specific sub-regions in the parameter space defined by the absolute value of the diurnal estimates and the value of the standard error estimate of the diurnal estimate. In contrast, criterion (3) does not strictly defines a sub-region in that parameter space, but rather an average region which can be visualized by mapping in that space the average number of data points used for the estimations. On average, estimates obtained with 24 data points or more are found in the parameter space for which the diurnal anomaly estimates are smaller than 10$^\circ$ and the standard error of the estimates are most often smaller than 1$^\circ$. In conclusion, we use the three criteria listed above to define self-exclusive quality flags as follows: a quality flag 5 indicates that all criteria (1), (2), and (3) are fulfilled; a quality flag 4 indicates that (1) and (2) are fulfilled but not (3); a quality flag 3 indicates that (1) is fulfilled but not (2) nor (3); and quality flag 2 indicates that none are fulfilled. {For analyses of diurnal SST, we recommend utilizing only estimates with quality flag 5. Further, a user may want to discard diurnal SST estimates for which the corresponding total and non-diurnal SST estimates both have quality flag less than 4. This occurs for less than 0.03\% of the diurnal SST estimates with quality flag 5.}

The inventory of {Level-3} estimates for each type (total, non-diurnal, and diurnal) and each quality flag class (0 to 5) is provided in Table~\ref{tab:quality}. The number of position and velocity estimates for the GDP hourly dataset version {2.00} is 165,754,333 from 17,324 individual drifter trajectories. Of this target number, 95.59\% are with a quality flag 5 for the total SST estimates, 95.60\% are with a quality flag 5 for the non-diurnal SST estimates, but only 75.58\% are with quality flag 5 for the diurnal SST estimates. Note that estimates of total SST and non-diurnal SST with quality flag 3 or 2 are outside the physically-acceptable range of values and should be used and interpreted with extreme caution. We assessed that diurnal SST estimates with quality flag 5 are plausible but we could not conclude the same for lesser quality flags.

\subsection*{Interpretation of uncertainty estimates}
\label{sec:interpr-uncert-estim}

In order to interpret our uncertainty estimates, we examine the distribution of the residuals of all model fits, normalized by their associated estimates of error standard deviations. This constitutes an assessment of the distribution of the error term $\varepsilon_i$ of the process model (\ref{eq:model0}):
\begin{linenomath*}\begin{equation}
\widehat{\varepsilon}_i = \frac{s_k -\widehat{s}_{m,k}}{\widehat{\sigma}(t_k)}.
\end{equation}\end{linenomath*}
The results are shown in Figure \ref{fig:residuals_spurs} for both the SPURS and test drifter subsets. For both sets, the distributions are never Gaussian for any of the models. The distributions are nearly centered but exhibit central peaks more narrow than Gaussian distributions with the same means and standard deviations (only comparisons to model 5 are shown). We observe that increasing the number of harmonics of the diurnal oscillation model consistently renders the peak of the residual distribution to be narrower and higher, and the tails to be slightly lighter. The opposite is true when increasing the order of the non-diurnal polynomial model, while still being non Gaussian in the sense of exhibiting a higher kurtosis. We find that a $t$ location-scale distribution (also known as non-standardized Student’s $t$ distribution), previously used to model Argos location errors \cite{elipot2016global}, is a better fit to the observed distributions than Gaussian distributions, yet still does not completely capture their shapes (not shown). An implication of the non-gaussianity of the normalized residuals is that the error term $\varepsilon_i$ of the process model (\ref{eq:model0}) is also not Gaussian-distributed. As a result, a classic least squares estimation of the parameters of the models would tend to give too much weight to outliers in the data. Fortunately, we are applying an iterative least squares estimation method based on the LOWESS \cite{Cleveland1979} which is expected to temper such outliers, but the exact impact on the estimation is difficult to quantify here. 

Nevertheless, a further implication of the non-gaussianity is that caution should be taken when interpreting the standard errors for the SST estimates described above: whereas for Gaussian-distributed errors one standard error can be used to calculate a 68\% confidence interval for an estimate, in our case, a standard error represents an interval encompassing more probable values of the true unknown values of a quantity, thus a more conservative confidence interval. As shown in Figure \ref{fig:residuals_spurs} for model 5, the 16-th and 84-th percentiles, encompassing 68\% of the residual distribution, define an interval narrower than the interval defined by plus or minus one sample standard deviation around the sample mean. Plus or minus one standard deviation actually encompasses approximately 78\% of the distribution of the residuals for model 5 (and approximately the same percentage for the other models, not shown). In other words, the standard error for our estimates can be interpreted as being representative of a 78\% confidence interval rather than a 68\% confidence interval. In contrast, the 2.5-th and 97.5-th percentiles, encompassing 95\% of the distribution, define an interval slightly wider but close to the one defined by plus or minus 1.96 sample standard deviation around the sample mean, which encompasses approximately 94\% of the distribution of the residuals for model 5 (and approximately the same percentage for the other models, not shown). In conclusion, considering 1.96 standard errors to quantify uncertainty in this case happens to represent an approximate 95\% confidence interval, as would be the case if the errors were Gaussian-distributed. Note that for the {Level-3} hourly product (Table \ref{tab:datamatrix}), the uncertainty estimates provided for location and velocity is 95\% confidence intervals, whereas for SST estimates the uncertainty estimates are standard error estimates.

\subsection*{Global characteristics of error variance estimates and uncertainty estimates}
\label{sec:glob-char-error}

%From Poli et al. 2019: In other terms, the SVP-BS data record confirms the ex- pectation that once the drogue is lost, the SST probes on a drifter are more likely to be exposed to water immediately below below the surface than when the drogue is present, and this effect is more visible in the presence of stratification (e.g., during daytime). To keep track of the drogue effect on SST measurements, it is important to monitor drogue loss as well the immersion depth and its variations.

In Figure \ref{fig:errorvar}, we examine the distribution of error variance estimates from residuals [$\widehat{\sigma}_1^2$, eq.~(\ref{eq:wrss1})] not including data points for which the error estimation failed, and the distribution of total error variance estimates incorporating the resolution error variance [$\widehat{\sigma}^2$, eq.~(\ref{eq:wrss3})] for {Level-3} estimates. We show the distributions for {Level-3} estimates only because the ones for {Level-2} estimates are extremely similar. We also report some statistics rounded to the nearest 0.001 in Table~\ref{tab:errors}, which differ by no more than 0.001$^\circ$C between {Level-2} and {Level-3} estimates. Based on the distributions in Figure~\ref{fig:errorvar}, we assess that the mode value, or most probable value, of the square root of the error variance estimates from residuals is 0.020$^\circ$C for drogued drifters, but is 50\% larger at 0.030$^\circ$C for undrogued drifters. Over all data, the mode value is 0.026$^\circ$C. Further, we assess that the mode value of the square root of the error variance estimates incorporating the resolution error variance is 0.031$^\circ$C for drogued drifters and 0.036$^\circ$C for undrogued drifters. Over all data, the mode value of the total error variance estimates is 0.033$^\circ$C. Median values of each of these variables are typically higher by a few 1/1000-th of a degree (See Table~\ref{tab:errors}): the overall median value of the square root of the total error variance estimates is 0.036$^\circ$C. The distribution of the total error variance is however not unimodal (Figure~\ref{fig:errorvar}, right) because of the resolution error variance is dominated by a few discrete values (Figure~\ref{fig:resolution}). 

The error variance estimates are however very heterogeneous in space, which is revealed when these estimates are averaged in half-degree geographical bins (Figure~\ref{fig:error_maps}, top). The spatial distribution of the error variance estimates is clearly related to ocean surface dynamics: it is found to be the highest in regions of high surface kinetic energy such as western boundary currents and equatorial regions \cite{laurindo2017improved}, but is also relatively high at mid-latitudes within regions of high wind stress variability. Largest mean error variance estimates are found on average within the Agulhas Retroflection region in the Indian Ocean and north of the Gulf Stream in the North Atlantic Ocean. The geographical distribution of error variance estimates suggests that the temporal evolution model (\ref{eq:model1}) might be improved by allowing the order of the polynomial $s_P$ to change spatially, in order to reduce the error variance estimates in these regions. Such a spatially-dependent model is beyond the scope of the present work but may be investigated in the future.

Whereas an error variance estimate provides a local quantification of the magnitude of the background noise, an uncertainty estimate for SST [eq.~(\ref{eq:varbeta})] provides a statistical characterization of the distance between a SST estimate and the true, but unknown, SST value. In Figure~\ref{fig:esst}, we examine the distributions of standard error estimates for the non-diurnal SST estimates, the diurnal SST estimates, and the total SST estimate for {Level-3} data, and we report overall statistics rounded to the nearest 0.001 in Table~\ref{tab:uncertainties}. The results for {Level-2} data are extremely similar and their distributions are not shown. Overall, the uncertainty estimates for diurnal SST estimates are a factor of 2 to 3 times larger than the uncertainty estimates for non-diurnal estimates. In turn, the uncertainty estimates for total SST estimates are larger than the uncertainty estimates for diurnal SST estimates but by no more than a few 1/1000-th of a degree. The most probable value of the uncertainty estimate for non-diurnal SST estimate is 0.006$^\circ$C for all data. For drogued drifters only it is 0.005$^\circ$C, and for undrogued drifters only it is 0.006$^\circ$C. The most probable value of the uncertainty estimate for diurnal SST estimate is 0.016$^\circ$C for all data. For drogued drifters only it is 0.015$^\circ$C, and for undrogued drifters only it is 0.017$^\circ$C. The most probable value of the uncertainty estimate for total SST estimate is 0.018$^\circ$C for all data. For drogued drifters only it is slightly smaller at 0.016$^\circ$C, and slightly higher for undrogued drifters at 0.019$^\circ$C. The spatial distribution of the uncertainty estimates follow closely the spatial distribution of the error variance estimates (Figure~\ref{fig:error_maps}, middle and bottom). 
%Locally, the average uncertainty estimates in half-degree bins for non-diurnal and diurnal SST estimates can be as high as x$^\circ$C and y$^\circ$C in western boundary region, respectively, and can be as low as x$^\circ$C and y$^\circ$C in subequatorial regions. 
The spatial distribution of the uncertainty estimates for total SST estimates is not shown as it is extremely similar to the spatial distribution of the uncertainty estimates for diurnal SST estimates. {The fact that the maps of averaged uncertainty estimates exhibit spatially coherent features related to geophysical variability implies that uncertainty estimates between drifters may also be correlated in relation to the geographical distance separating them.}

The overall statistics of uncertainty estimates for SST estimates (Table~\ref{tab:uncertainties}) are an order of magnitude smaller than previously estimated measurement uncertainties for drifting buoys \cite{kennedy2014review}. Such uncertainty estimates range between 0.1 $^\circ$C and 0.7 $^\circ$C and are typically based on analyses of collocated SST observations from drifting buoys, ships, and satellites \cite{emery2001accuracy,ocarroll2008threeway,xu2010evaluation,merchant2012a20year}. These uncertainty estimates encompass not only the instrumental error of the drifter SST sensors but also the spatial and temporal differences between the different measurands that are targeted by the different observational platforms, such as a SST satellite’s ground footprint versus a pointwise drifter measurement. {Here, our uncertainty estimates represent only random sources of instrumental and communication noise, as well as sub-hourly unresolved geophysical variability. These uncertainties are on the order of 1/100-th of a Kelvin, rather than on the order of a 1/10-th of a Kelvin, because they are not estimated from a single observation, but rather are benefiting from time series of observations that typically provide around 22 effective degrees of freedom over a 2-day observational estimation window (see mode value of $\nu$ in Figure} \ref{fig:deno_errorvar}{). As a result, sources of instrumental and geophysical random noise are averaged downward for our estimates. What our uncertainty estimates are not able to capture is any drifter-specific original bias of a SST sensor, and which may, or not, have evolved in time since the times of manufacture and deployment (i.e. a sensor drift)}\cite{poli2019copernicus}.

% \subsection*{Subsection}
% Example text under a subsection. Bulleted lists may be used where appropriate, e.g.
% \begin{itemize}
% \item First item
% \item Second item
% \end{itemize}
% \subsubsection*{Third-level section}
%  Topical subheadings are allowed.

\section*{Data Records}

%The Data Records section should be used to explain each data record associated with this work, including the repository where this information is stored, and to provide an overview of the data files and their formats. Each external data record should be cited numerically in the text of this section, for example \cite{Hao:gidmaps:2014}, and included in the main reference list as described below. A data citation should also be placed in the subsection of the Methods containing the data-collection or analytical procedure(s) used to derive the corresponding record. Providing a direct link to the dataset may also be helpful to readers (\hyperlink{https://doi.org/10.6084/m9.figshare.853801}{https://doi.org/10.6084/m9.figshare.853801}).
%Tables should be used to support the data records, and should clearly indicate the samples and subjects (study inputs), their provenance, and the experimental manipulations performed on each (please see 'Tables' below). They should also specify the data output resulting from each data-collection or analytical step, should these form part of the archived record.

The {Level-3} estimates of total SST, non-diurnal SST, diurnal SST, and each of their respective standard error estimates, along with quality flag variables for each of the three SST estimates, are distributed as part of the hourly drifter dataset of the GDP \cite{elipot2016global}, now in its version {2.00} with the addition of these SST estimates. The dataset, assembled as a contiguous ragged array in a single file, is officially available from the NOAA National Center for Environmental Information (NCEI) {as a data collection}\cite{elipot2022hourly} {called "Hourly location, current velocity, and temperature collected from Global Drifter Program drifters world-wide'' and accessible at} \url{https://doi.org/10.25921/x46c-3620}. The original and future releases (also called accession) of this collection can be accessed and downloaded through the ``Lineage'' tab of the landing page. Future releases of the dataset, scheduled twice a year, will add estimates of position, velocity, and SST variables as they become available from the GDP.

The data are also available via the ERDDAP server of the NOAA Observing System Monitoring Center at \url{http://osmc.noaa.gov/erddap/tabledap/gdp\_hourly\_velocities.html} where subsets of the data can be selected according to a number of temporal and spatial criteria.
%In addition, individual NetCDF files, one per drifter, are available for batch download via the secure FTP server of the Data Assembly Center (DAC) at the NOAA Atlantic Oceanographic and Meteorological Laboratory at \url{ftp://ftp.aoml.noaa.gov/pub/phod/lumpkin/hourly/v2.0/netcdf/}. The names of the files are \texttt{drifter\_X.nc} where \texttt{X} is the AOML ID number for Argos drifters and the International Mobile Equipment Identity (IMEI) number for Iridium drifters. 

Table~\ref{tab:detail} lists the names of the variables included in the NetCDF files, including the new SST-related variables. Usage of this SST data product in combination with any of the position and/or velocity data \cite{elipot2016global} for release 2.00 or subsequent releases must cite this present paper as well as the original 2016 paper describing the hourly position and velocity dataset (Elipot et al. 2016\cite{elipot2016global}). 

\section*{Technical Validation}

%\emph{This section presents any experiments or analyses that are needed to support the technical quality of the dataset. This section may be supported by figures and tables, as needed. This is a required section; authors must present information justifying the reliability of their data.}

The spatial and temporal distributions of the {Level-3} hourly SST estimates are displayed in Figure~\ref{fig:density_histogram} in order to verify the technical quality of the product. The map of spatial data density is the result of historical deployments and the efforts of the GDP to fulfill the requirement of the array, and of the patterns of the convergence and divergence of the near-surface oceanic circulation \cite{lumpkin2016fulfilling,lumpkin2012evaluating}. The temporal histogram of SST estimates closely follows the distribution of hourly position and velocity estimates, showing the maturity of the array at the beginning of 2006 as well as the drop in the amount of data between 2011 and 2014 because of unfortunately numerous short-lived instruments. To support the technical validation of the new SST dataset, we compute the mean and standard deviation of SST estimates globally within $0.5^\circ \times 0.5^\circ$ geographical bins (Figure~\ref{fig:mean_std}). The mean total SST map exhibits the expected meridional gradients as well as the west-east asymmetries within each ocean basins. As also expected, the standard deviation map of total SST estimates exhibits larger values within regions of higher surface kinetic energy such as in western boundary current regions \cite{lumpkin2013global} but also within the mid-latitude regions where high variability of air-sea fluxes is expected to enhance SST variance. The map of diurnal SST standard deviation exhibits different patterns resulting from the competing effects of the spatial pattern of solar heating increasing diurnal variability, and the spatial pattern of wind speed decreasing diurnal variability. At the scales displayed here, the maps of mean and standard deviation of non-diurnal SST estimates (not shown) are indistinguishable from the maps for the total SST estimates. The map of mean diurnal SST estimates (not shown) is approximately zero everywhere as expected from the model of temporal SST evolution used to derive this product.

We proceed to verify the consistency of the drifter hourly SST estimates against the gridded, multi-sensor, interpolated SST Climate Change Initiative data product (ESA SST CCI Analysis v2.1, hereafter CCI), available from 1981 to 2016\cite{good2019esa}. The CCI product is generated by combining measurements of infrared radiance from two suites of radiometers on multiple satellites, eventually aggregated and gap-filled on a daily 0.05$^\circ$grid \cite{merchant2019satellite}. The CCI product provides SST estimates representative of daily mean values and at a depth of 20 cm. These estimates are obtained by converting the instantaneous skin SST captured by satellite measurements at various times throughout a day to the closest of 10:30 or 22:20 local mean solar times, using a one-dimensional turbulence closure model driven by atmospheric fluxes. Arguably, this conversion makes the CCI estimates comparable to the drifter SST estimates because of the depth conversion, yet differences are expected to remain between the two estimates because of the drifter SST sub-daily temporal variability mostly associated with the diurnal cycle. We conduct the comparison by interpolating bilinearly in space the gridded values of the CCI product for a given day onto all drifter hourly geographical locations of that same day (from 00:00 to 23:00). SST estimates derived from satellite measurements are not independent from in situ measurements, including from drifters, because these are generally used for calibration purposes. Yet, the ESA SST CCI Analysis v2.1 is supposed to achieve a ``high degree of independence'' from in situ observations from 1995 onward\cite{merchant2019satellite}, so that we limit our comparison to the 1995 to 2016 time period. We choose to compare only drifter total and non-diurnal SST estimates with quality flag 5 to successfully interpolated values of the CCI product, that is when no land pixel or pixel with non-zero sea ice concentration were involved in the bilinear interpolation. Two-dimensional histograms of drifter SST estimates versus their corresponding CCI interpolated values for nearly 122M data pairs (Figure\ref{fig:sst_vs_cci}a,b) suggest qualitatively a very good consistency between the two datasets. Next, we examine difference statistics between the two datasets in order to conduct a more quantitative comparison. We interpolate the evaluated standard uncertainty of the CCI product at the drifter locations in order to assess the difference statistics.

We calculate the difference statistics $d_m = \widehat{s}_m - s_{cci}$ and $d_P = \widehat{s}_P - s_{cci}$, where $s_{cci}$ is the interpolated CCI SST value, and the uncertainties corresponding to these differences, i.e. $(\widehat{\sigma}^2_m+\sigma_{cci}^2)^{1/2}$ and $(\widehat{\sigma}^2_P+\sigma_{cci}^2)^{1/2}$, where $\sigma_{cci}$ is the interpolated CCI SST uncertainty value. Over all estimates, we find that the mean and median values of $d_m$ are 0.048$^\circ$C and 0.031$^\circ$C, respectively, and that the mean and median values of $d_m$ are 0.047$^\circ$C and 0.037$^\circ$C, respectively, suggesting a global positive bias of the drifter estimates compared to the CCI estimates. We next conduct a 95\% confidence level two-tailed test by determining the number of instances for which the absolute values of the difference statistics are smaller than 1.96 times the difference uncertainties. We thus find that 85.5\% of the differences with the drifter total SST estimates are not statistically significant, whereas we find that 88.1\% of the differences with the drifter non-diurnal SST estimates are not significant, overall suggesting a high level of consistency between the two products. Because the CCI product does not resolve diurnal variability, it is expected that the consistency between the two datasets would be better for the drifter non-diurnal estimates than for the drifter total estimates, as evidenced here by the two-tailed test results. However, the proportion of statistically significant differences for the non-diurnal drifter SST estimates is still over twice the expectation from a 95\% confidence level test, at 11.9\% compared to 5\%. This excess proportion may be due to the incorrect assumption of Gaussian-distributed errors made in our two-tailed test where heavier tailed errors are expected in practice (see section \emph{Interpretation of uncertainty estimates} and Figure~\ref{fig:errorvar}), stochastic variability unresolved by the uncertainty estimates of either data product, loss of spatial resolution of the CCI analysis product because of its mapping, or a global bias as revealed by the mean difference values reported above. The most probable difference of values between the two products, as indicated by the maximum of the distribution of the absolute difference statistics (Figure~\ref{fig:sst_vs_cci}c) is 0.250$^\circ$C with the drifter total SST estimates, and 0.223$^\circ$C with the drifter non-diurnal SST estimates. These mode values are roughly consistent with a linear sum of the stated global uncertainty for the CCI product\cite{merchant2019satellite}(0.18$^\circ$C), plus the typical uncertainty values of the drifter estimates (0.018$^\circ$C and 0.007$^\circ$C, see Table~\ref{tab:uncertainties}), plus potential global biases (0.048$^\circ$C and 0.047$^\circ$C), amounting to 0.246$^\circ$C and 0.234$^\circ$C for the total SST estimates and the non-diurnal SST estimates, respectively.

While it is beyond the scope of this paper to systematically investigate the potential sources of differences between our drifter hourly SST dataset and the CCI product (or other satellite-based products), we further examine the differences between the drifter non-diurnal SST estimates and the CCI values as a function of latitude and time in 10-day intervals from 1996 to 2016 (Figure\ref{fig:sst_vs_cci}d). This analysis is consistent with a similar analysis conducted previously\cite{merchant2019satellite}, but provides more details and is performed with our Level 4 hourly drifter product. Our results show that the drifter non-diurnal estimates generally exhibit a positive bias compared to the CCI product, except between approximately 15$^\circ$ and 45$^\circ$, N or S, where the differences exhibit alternating signs with an annual periodicity propagating poleward. These propagating differences might be related to inadequacies of the turbulent closure model used to convert satellite-measured skin SST to CCI depth SST, or to the inability of representing accurately seasonal processes in either the model or the forcing fields of the model. We also examine the dependency of differences on local mean solar time (Figure~\ref{fig:sst_minus_cci_lst}). We find that the differences between drifter total SST estimates and CCI values as a function of local mean solar time follow a distribution pattern consistent with the typical SST diurnal cycle\cite{morakbozzo2016climatological}: the drifter total SST estimates generally capture a higher temperature between the hours of 10:30 and 22:30, but generally capture a lower temperature between 22:30 and 10:30 the next day (panel a). In contrast, the distribution of differences between the drifter non-diurnal SST estimates and CCI values do not appreciably exhibit a dependency on local mean solar time, as expected, but only an overall positive bias (panel b).

%To further check these data, we computed global maps of average SST and SST variance in half-degree bins over the World's Ocean using quality flag 5 data only (Figure~\ref{fig:mean_std}). The results show the expected meridional gradient for the map of mean SST, and the expected distribution of SST variance influenced by the oceanic general circulation and air-sea fluxes spatial distribution.

\section*{Usage Notes}

%\emph{The Usage Notes should contain brief instructions to assist other researchers with reuse of the data. This may include discussion of software packages that are suitable for analysing the assay data files, suggested downstream processing steps (e.g. normalization, etc.), or tips for integrating or comparing the data records with other datasets. Authors are encouraged to provide code, programs or data-processing workflows if they may help others understand or use the data. Please see our code availability policy for advice on supplying custom code alongside Data Descriptor manuscripts.}

In the NetCDF file, all SST estimates are provided to three decimal places, with the last digit rounded towards the nearest 0.001. Total SST estimates are the sum of non-diurnal SST estimates and diurnal SST estimates but because of rounding, discrepancies exist within the NetCDF files for about 44\% of values between the numerical value the user will read for the total SST value and the value of the sum of the non-diurnal and diurnal SST values.

The uncertainty estimates are also provided to three decimal places but with the last digit rounded ``up'' (i.e. towards infinity) to the nearest 0.001. The reason for the rounding up of the uncertainty estimates is to prevent reporting null uncertainties for 13,502 non-diurnal SST estimates for which the calculated uncertainty is smaller than 0.001 but larger than 0.0001. Rounding up uncertainties is acceptable as this provides more conservative uncertainties but only increasing their values by typically 6\% for the non-diurnal SST uncertainties, and by typically 2\% for the diurnal and total SST uncertainties.

{Instructions on how to read the data file, as well as examples of typical Lagrangian analyses, can be found in a publicly-accessible Python-based Jupyter Notebook at} \url{https://github.com/Cloud-Drift/earthcube-meeting-2022}.

%For studies involving privacy or safety controls on public access to the data, this section should describe in detail these controls, including how authors can apply to access the data, what criteria will be used to determine who may access the data, and any limitations on data use. 

\section*{Code availability}

% For all studies using custom code in the generation or processing of datasets, a statement must be included under the heading "Code availability", indicating whether and how the code can be accessed, including any restrictions to access. This section should also include information on the versions of any software used, if relevant, and any specific variables or parameters used to generate, test, or process the current dataset. 

%Part of the software (version 1.0.0) associated with this manuscript for the calculation and storage of PSDs is licensed under MIT and published on GitHub https://github.com/Jollyfant/psd-module/ (Jollyfant, 2021).

A software associated with this manuscript is licensed under MIT and published on GitHub at \url{https://github.com/selipot/sst-drift.git} and archived on Zenodo \cite{elipot2021sstdrift}. This software allows the user to fit model (\ref{eq:model1}) to temperature observations and derive the resulting SST estimates and their uncertainties. Input arguments to the model fitting function include an arbitrary order for the background non-diurnal SST model and arbitrary frequencies for the diurnal oscillatory model. A sample of {Level-1} data from drifter AOML ID 55366 is provided in order to test the routines and produce figures similar to Figures~\ref{fig:example1} and \ref{fig:example2}. Alternatively, the main code can also generate stochastic data for testing purposes. 

%\noindent LaTeX formats citations and references automatically using the bibliography records in your .bib file, which you can edit via the project menu. Use the cite command for an inline citation, e.g. \cite{Kaufman2020, Figueredo:2009dg, Babichev2002, behringer2014manipulating}. For data citations of datasets uploaded to e.g. \emph{figshare}, please use the \verb|howpublished| option in the bib entry to specify the platform and the link, as in the \verb|Hao:gidmaps:2014| example in the sample bibliography file. For journal articles, DOIs should be included for works in press that do not yet have volume or page numbers. For other journal articles, DOIs should be included uniformly for all articles or not at all. We recommend that you encode all DOIs in your bibtex database as full URLs, e.g. https://doi.org/10.1007/s12110-009-9068-2.

\section*{Acknowledgements} 

%Acknowledgements should be brief, and should not include thanks to anonymous referees and editors, or effusive comments. Grant or contribution numbers may be acknowledged.

This research was partially carried out under the auspices of the Cooperative Institute for Marine and Atmospheric Studies (CIMAS), a Cooperative Institute of the University of Miami and the National Oceanic and Atmospheric Administration, cooperative agreement \#NA20OAR4320472. This research was also supported by the US National Science Foundation under EarthCube Capabilities Grant No. 2126413. A. M. Sykulski was funded by the UK Engineering and Physical Sciences Research Council Grant EP/R01860X/1. R. Lumpkin was supported by NOAA's Global Ocean and Monitoring Program and the Atlantic Oceanographic and Meteorological Laboratory. L. Centurioni was supported by NOAA's grant NA20OAR4320278 ``The Global Drifter program''. The authors thank Sofia Olhede for her advice on some of the statistical aspects of this work, and Bertrand Dano and Philippe Miron for generating the final NetCDF file for distribution.

\section*{Author contributions statement}

S. Elipot, A. Sykulski, and R. Lumpkin conceived the dataset and designed the associated methods. M. Pazos assembled and conducted the quality control of the drifter data forming the basis for the {Level-1} data. All authors reviewed the manuscript.

\section*{Competing interests} (mandatory statement)

All authors declare no conflict of interest.

%\bibliography{sample}
%\bibliography{/Users/selipot/Work/papers/SST.bib,references_added.bib}

\section*{Figures \& Tables}

\begin{figure}[ht]
\centering
\includegraphics[height=\textheight]{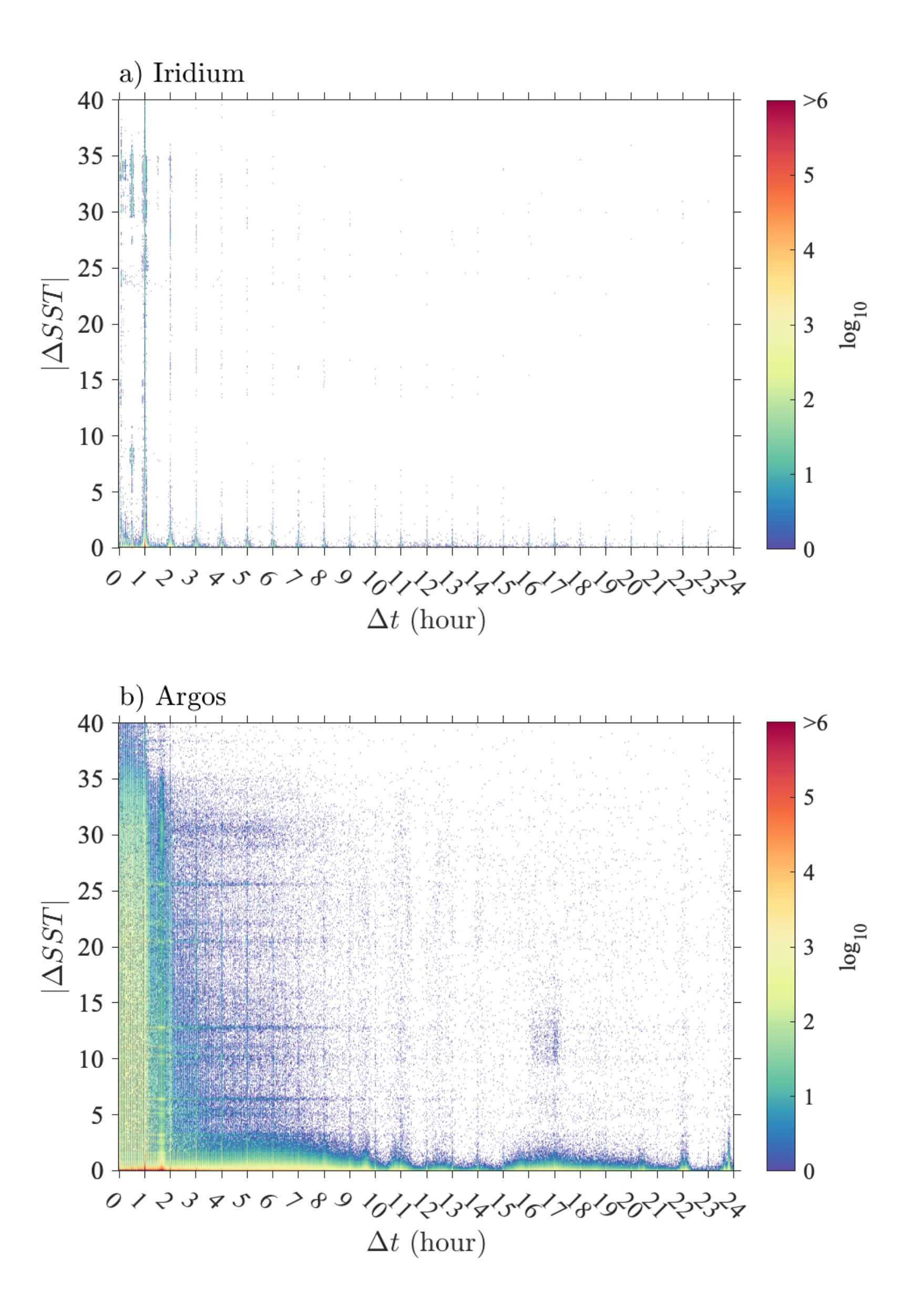}
\caption{Distribution of time differences and absolute temperature differences between two consecutive SST observations of the {Level-1} data. (a) Two-dimensional histogram for Iridium drifters, and (b) Two-dimensional histogram for Argos drifters. Only values of $\Delta$t less than 24 hours and $40^\circ$C are shown.}
\label{fig:deltasst_deltat}
\end{figure}

% 2
\begin{figure}[ht]
\centering
\includegraphics[width=\textwidth]{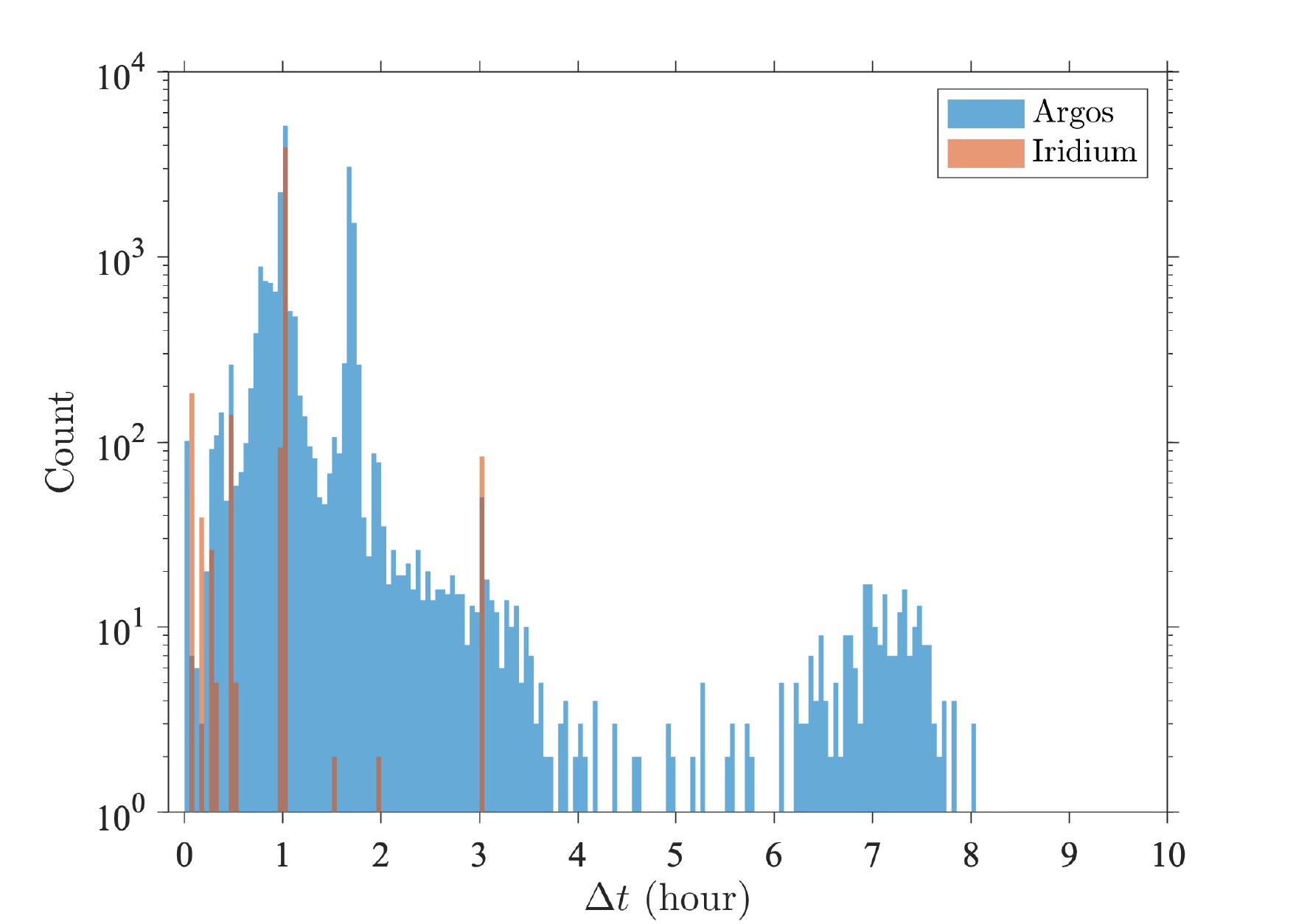}
\caption{Histograms in 3-minute bins of median SST temporal sampling intervals per drifter trajectory for {Level-1} data (24,597 SST time series from 4,495 Iridium drifters and 20,102 Argos drifters). Only values smaller than 10 hours are displayed. 105 Argos time series have a median sampling larger than 10 hours.}
\label{fig:deltat}
\end{figure}

% 3
\begin{figure}[ht]
\centering
\includegraphics[width=\textwidth]{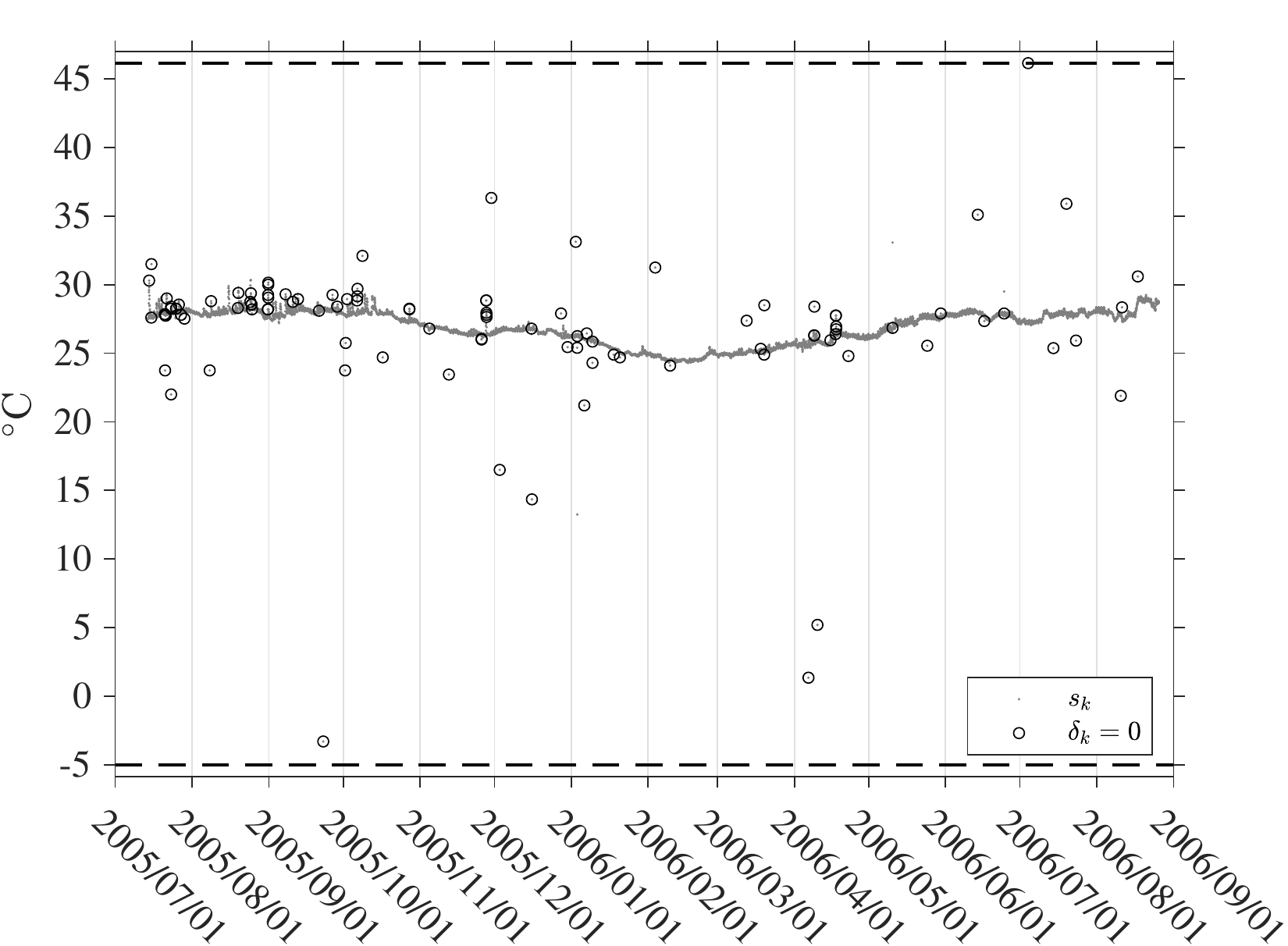}
\caption{Time series of SST data for GDP drifter ID 55366 (WMO number 3100541). This drifter was built by Pacific Gyre and is of the Surface velocity Program (SVP) type, tracked by the Argos positioning system. The median time interval between SST observations for this drifter is about 52 min. The SST equation for this drifter is SST($^\circ$C) = 0.05 $\times n$ - 5.00 where $n$ is a 10-bit sensor count. This equation defines the data resolution ($0.05^\circ$C) as well as the minimum value ($0.05 \times 0 -5.00 =  -5.00^\circ$C) and maximum value ($0.05 \times (2^{10}-1)-5.00 = 46.15^\circ$C) that should be returned by the temperature sensor, indicated by the horizontal dashed lines on this figure. The {Level-1} data ($s_k$) are indicated by gray dots. The circled dots are the data points that are ultimately down-weighed to zero by the iterative estimation method, and thus flagged as outliers ($\delta_k=0$).}
\label{fig:example0}
\end{figure}

% 4
\begin{figure}[ht]
\includegraphics[width=\textwidth]{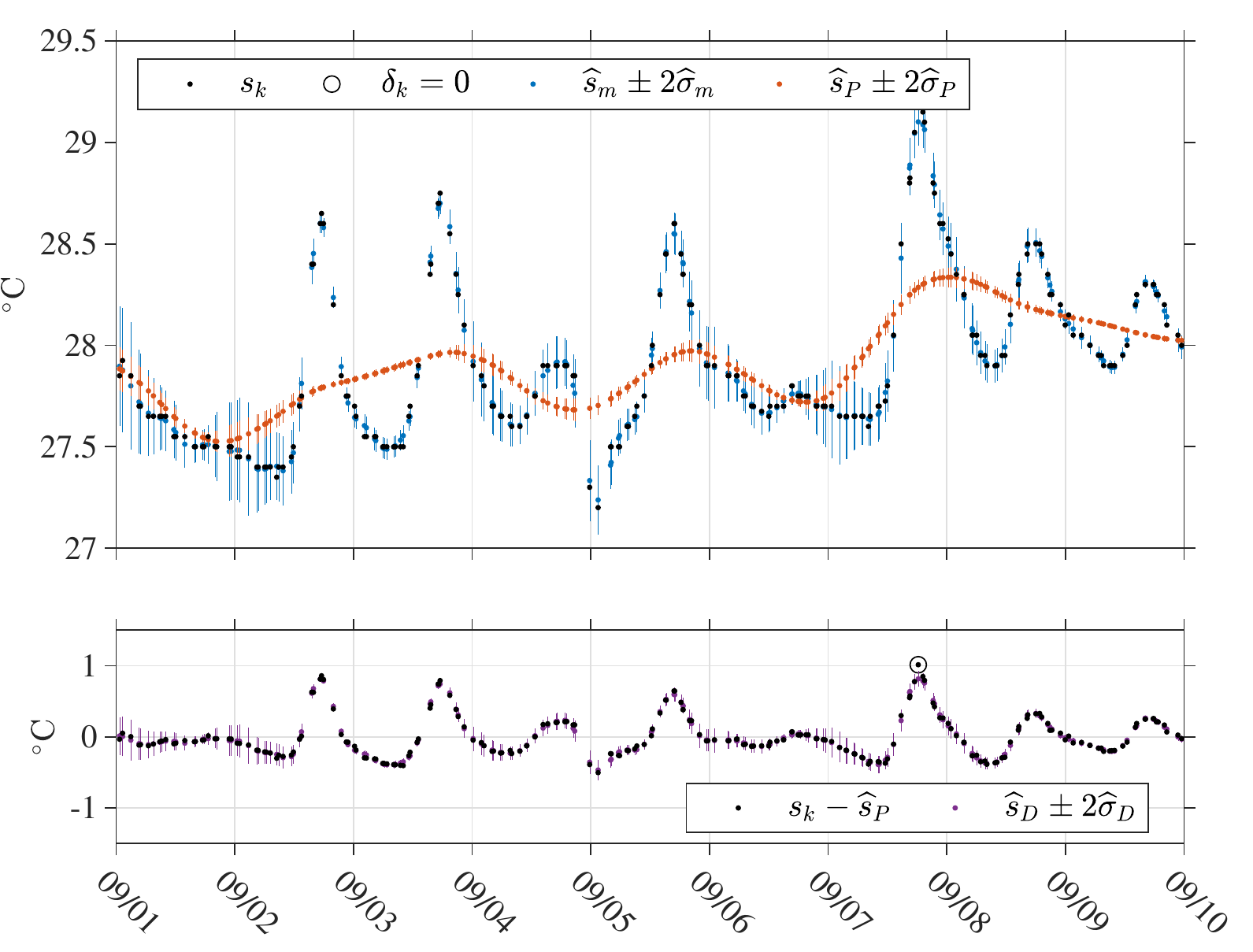}
\caption{Time series of SST estimates for GDP drifter ID 55366 (WMO number 3100541) between 2005/9/1 and 2005/9/10. Top panel: Black dots are the original SST data ({Level-1}, $s_k$) and circles are the data points down-weighed to zero ($\delta_k=0$). The blue dots with vertical lines show the total SST estimates and their plus or minus two standard errors ($\widehat{s}_m \pm 2\widehat{\sigma}_m$). The red dots and vertical lines show the non-diurnal SST estimates and their plus or minus two standard errors ($\widehat{s}_P \pm 2\widehat{\sigma}_P$). The blue estimates are the sum of the red estimates and purple estimates shown in the lower panel of the figure. Lower panel: Black dots show SST data minus the non-diurnal SST estimate  ($s_k - \widehat{s}_P$). The purple dots and vertical lines show the corresponding diurnal SST estimates and their plus or minus two standard errors ($\widehat{s}_D \pm 2\widehat{\sigma}_D$). }
\label{fig:example1}
\end{figure}

% 5
\begin{figure}[ht]
\includegraphics[width=\textwidth]{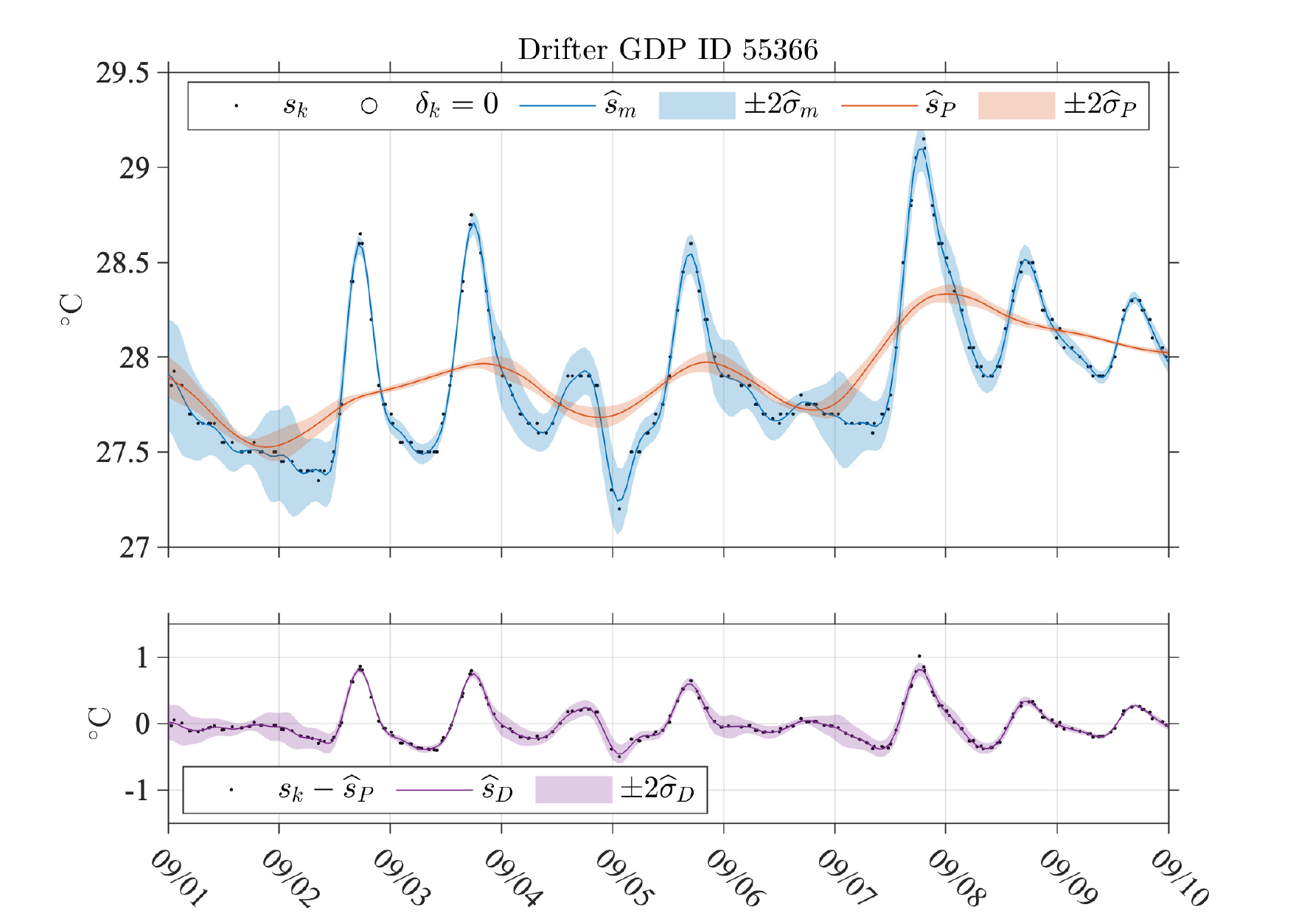}
\caption{Time series of continuous hourly SST estimates for GDP drifter ID 55366 (WMO number 3100541) between 2005/9/1 and 2005/9/10. Top panel: Black dots show the original SST data ({Level-1}, $s_k$). The blue line and shaded region show continuously the hourly total SST estimates and twice their standard errors ($\widehat{s}_m \pm 2 \widehat{\sigma}_m$). The red line and shaded region show continuously the hourly non-diurnal SST estimates and twice their standard errors ($\widehat{s}_m \pm 2 \widehat{\sigma}_P$). The blue line is the sum of the red line and the purple line shown in the lower panel. Lower panel: Black dots show SST data minus the non-diurnal SST estimate  ($s_k - \widehat{s}_P$). The purple line and shaded region show continuously the hourly diurnal SST estimates and twice their standard errors ($\widehat{s}_D \pm 2 \widehat{\sigma}_D$). }
\label{fig:example2}
\end{figure}

% 6
\begin{figure}[ht]
\centering
\includegraphics[width=\textwidth]{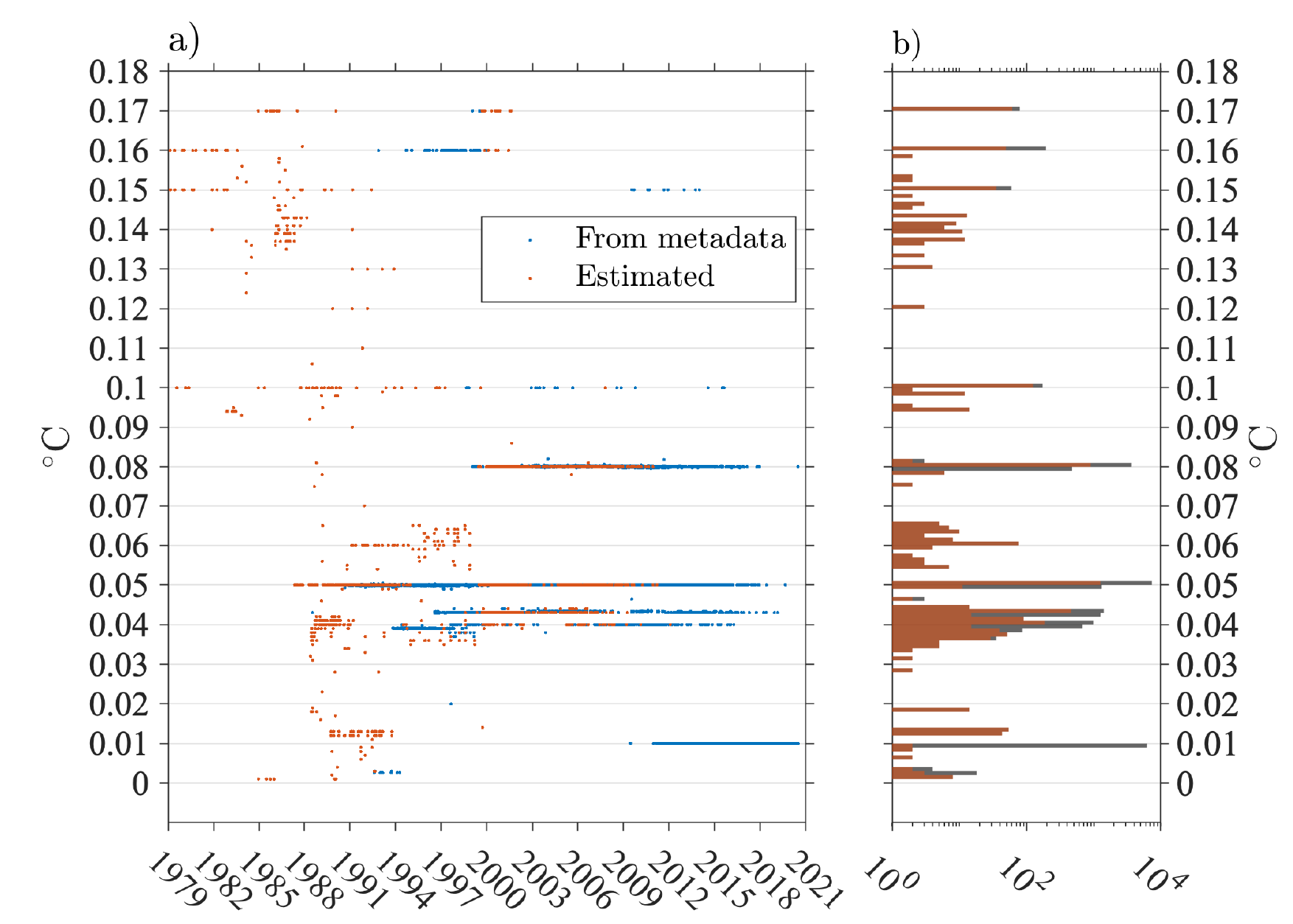}
\caption{Distributions of drifter SST sensors resolution.  a) Temporal distribution of drifter SST resolution in the GDP database from February 1979 to July 2020. The blue points corresponds to resolution $a$ obtained from the drifter metadata from the SST equation: SST($^\circ$C) =  $a \times n + b$ where $n$ is a bit sensor count. The red points corresponds to drifters for which the resolution is not available from the metadata and was estimated directly from the data (see text). b) Histogram of drifter SST resolution values in 0.001$^\circ$C bins. The red bars correspond to the estimated resolution values and the gray bars correspond to all values. Note that the horizontal axis in on a log scale. The three most common resolution values in the dataset are in order: 0.05, 0.01, and 0.08$^\circ$C.}
\label{fig:resolution}
\end{figure}

% 7
\begin{figure}[ht]
\centering
\includegraphics[width=\textwidth]{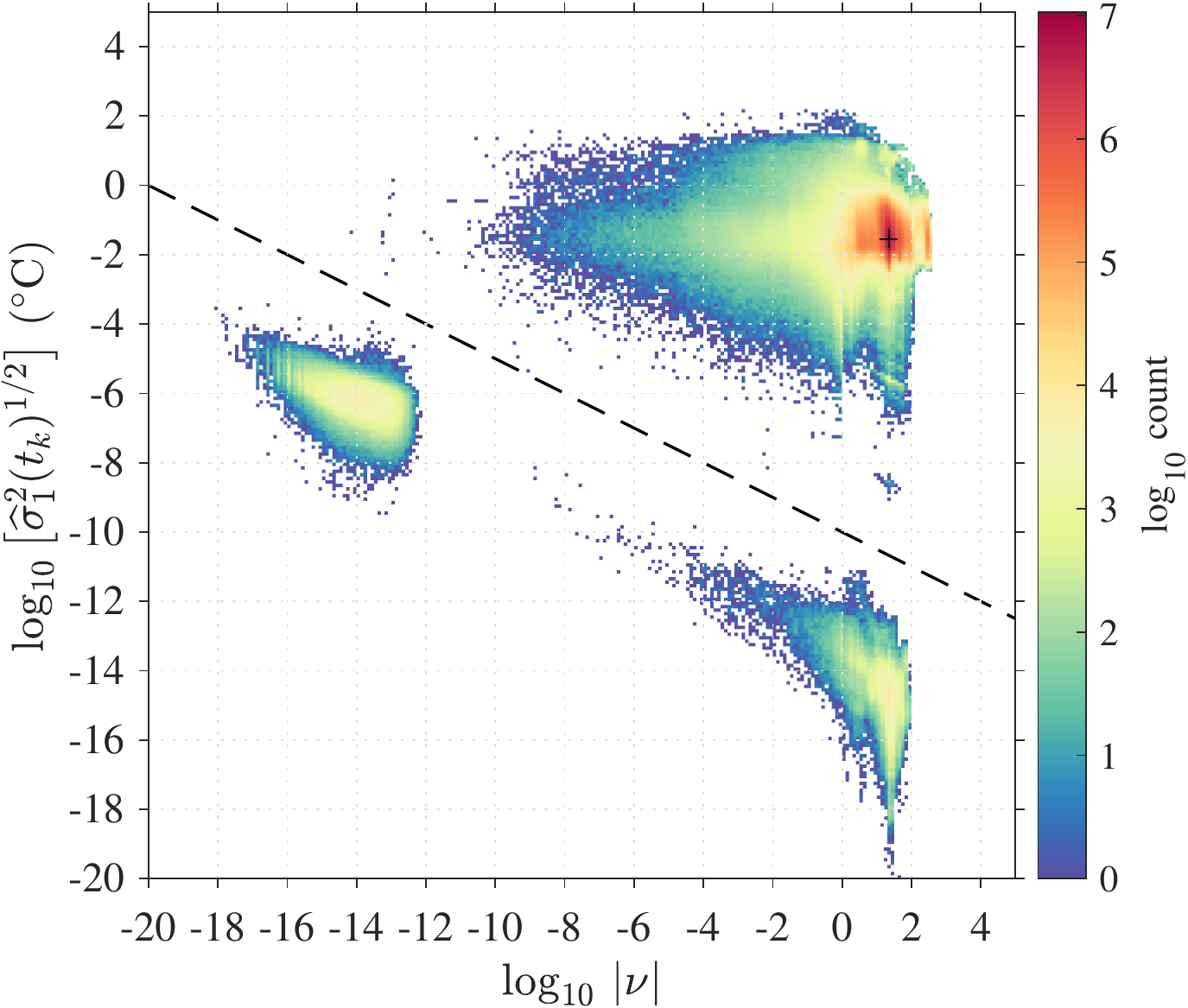}
\caption{Two-dimensional histogram of the effective degrees of freedom for the residuals ($\nu$) and estimates of the error variance neglecting the resolution error variance [$\widehat{\sigma}_1^2(t_k)$] for {Level-2} data results. The two populations found below the black dashed line ($\log_{10} [\widehat{\sigma}_1^2]^{1/2} < -\frac{1}{2}\log_{10} |\nu| -10$) correspond to failed estimations of the error variance and are flagged with quality flag 1. The left population below the dashed line (0.33\% of the data) corresponds to negative estimated variance (see text). The right population below the dashed line (0.22\% of the data) corresponds to uncharacteristically flat SST records leading to unrealistic near-zero estimated error variances. The peak of the distribution is found within the upper-right population for $\nu \approx 22.4$ and $[\widehat{\sigma}_1^2(t_k)]^{1/2} \approx 0.028^\circ$C.}
\label{fig:deno_errorvar}
\end{figure}

% 8
\begin{figure}[ht]
\includegraphics[width=\textwidth]{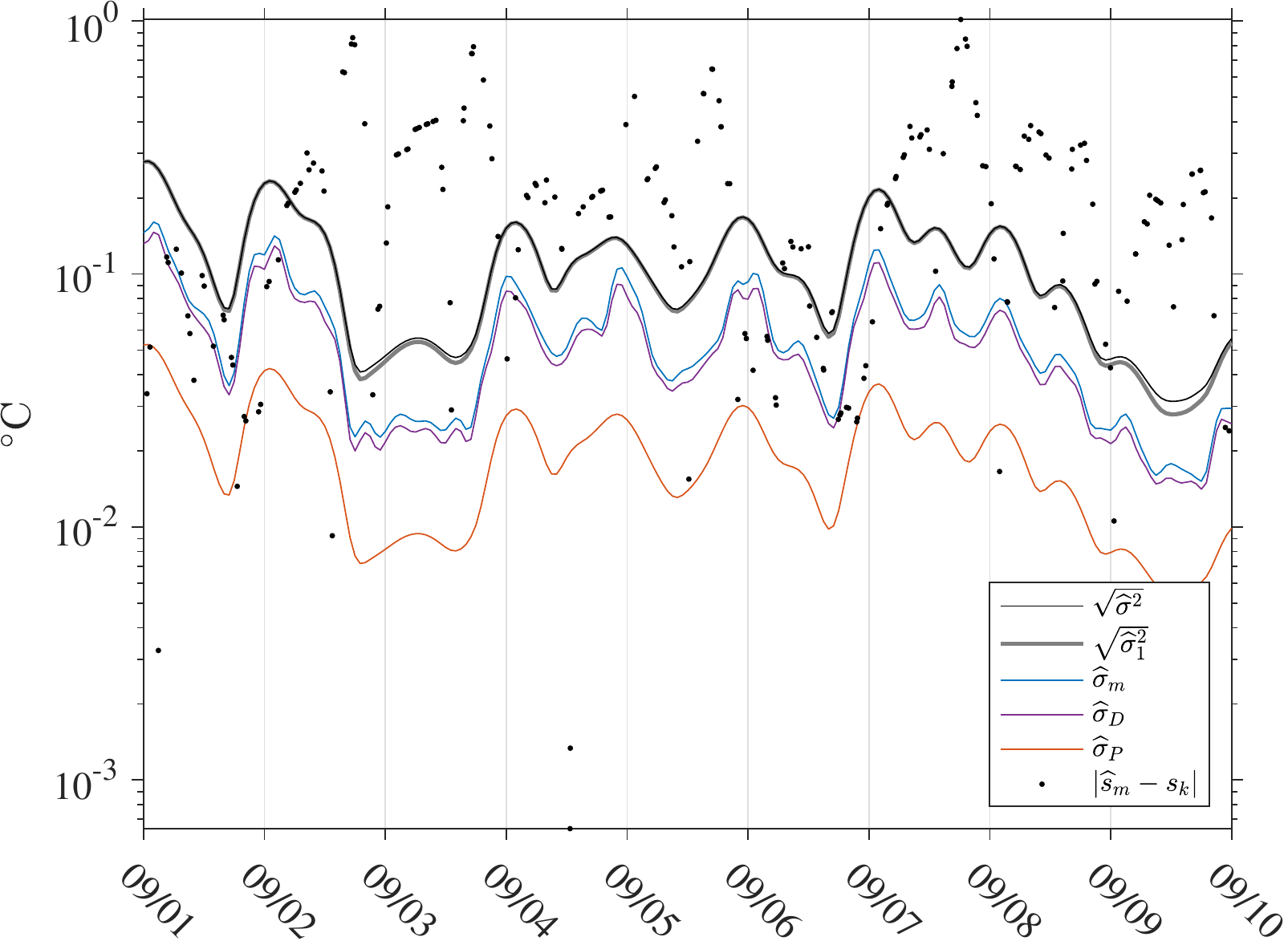}
\caption{Time series for GDP drifter ID 55366 (WMO ID 3100541) between 2005/9/1 and 2005/9/10 of SST standard error estimates ($\widehat{\sigma}_m$), non-diurnal SST standard error estimates ($\widehat{\sigma}_P$), diurnal SST standard error estimates ($\widehat{\sigma}_D$), square root of error variance estimates from residuals [$\sqrt{\widehat{\sigma}_1^2}$, eq.~(\ref{eq:wrss1})], square root of total error variance estimates [$\sqrt{\widehat{\sigma}^2}$, eq.~(\ref{eq:wrss3})], and absolute residuals ($|\widehat{s}_m-s_k|$). The curves for $\sqrt{\widehat{\sigma}_1^2}$ and $\sqrt{\widehat{\sigma}^2}$ are most often indistinguishable except around 09/03 and 09/09-09/10.}
\label{fig:example3}
\end{figure}

% 9
\begin{figure}[ht]
\includegraphics[width=\textwidth]{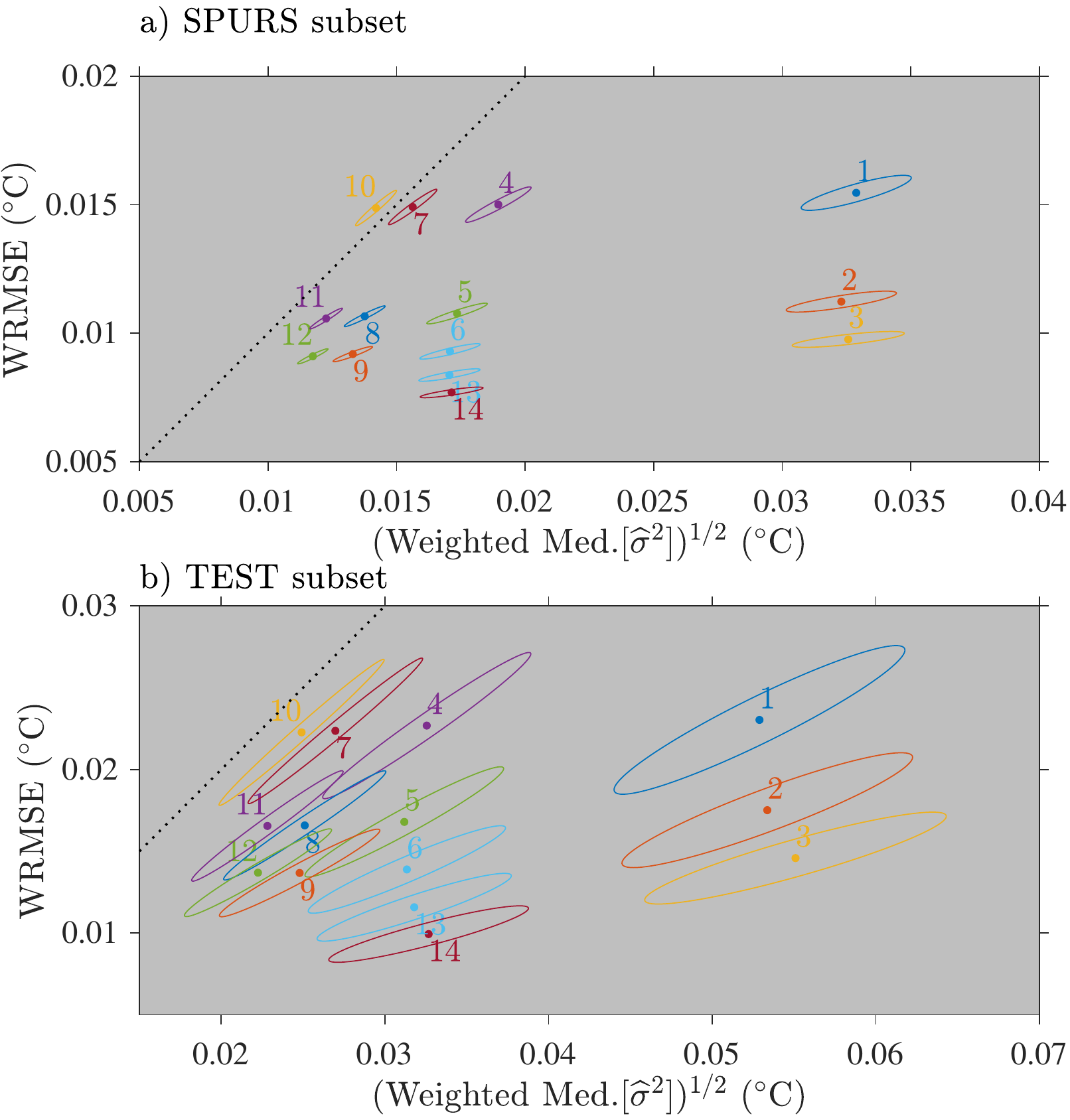}
\caption{Summary statistics for the 14 models listed in Table \ref{tab:models} for (a) the subset of 80 drifters from the SPURS experiment and (b) the subset of 98 ``test'' drifters selected from the global database. Colored dots with numbers indicate the average values of the square root of the weighted median of the error variance (horizontal axis) versus the average values of the weighted root mean square error (WRMSE, vertical axis). Ellipses correspond to 95\% confidence intervals across ensemble statistics. Note the different axis ranges between panels a) and b). The black dotted line indicates the slope-1 intercept-0 curve.}
\label{fig:modelperfs}
\end{figure}

% 10
\begin{figure}[ht]
\centering
\includegraphics[height=0.9\textheight]{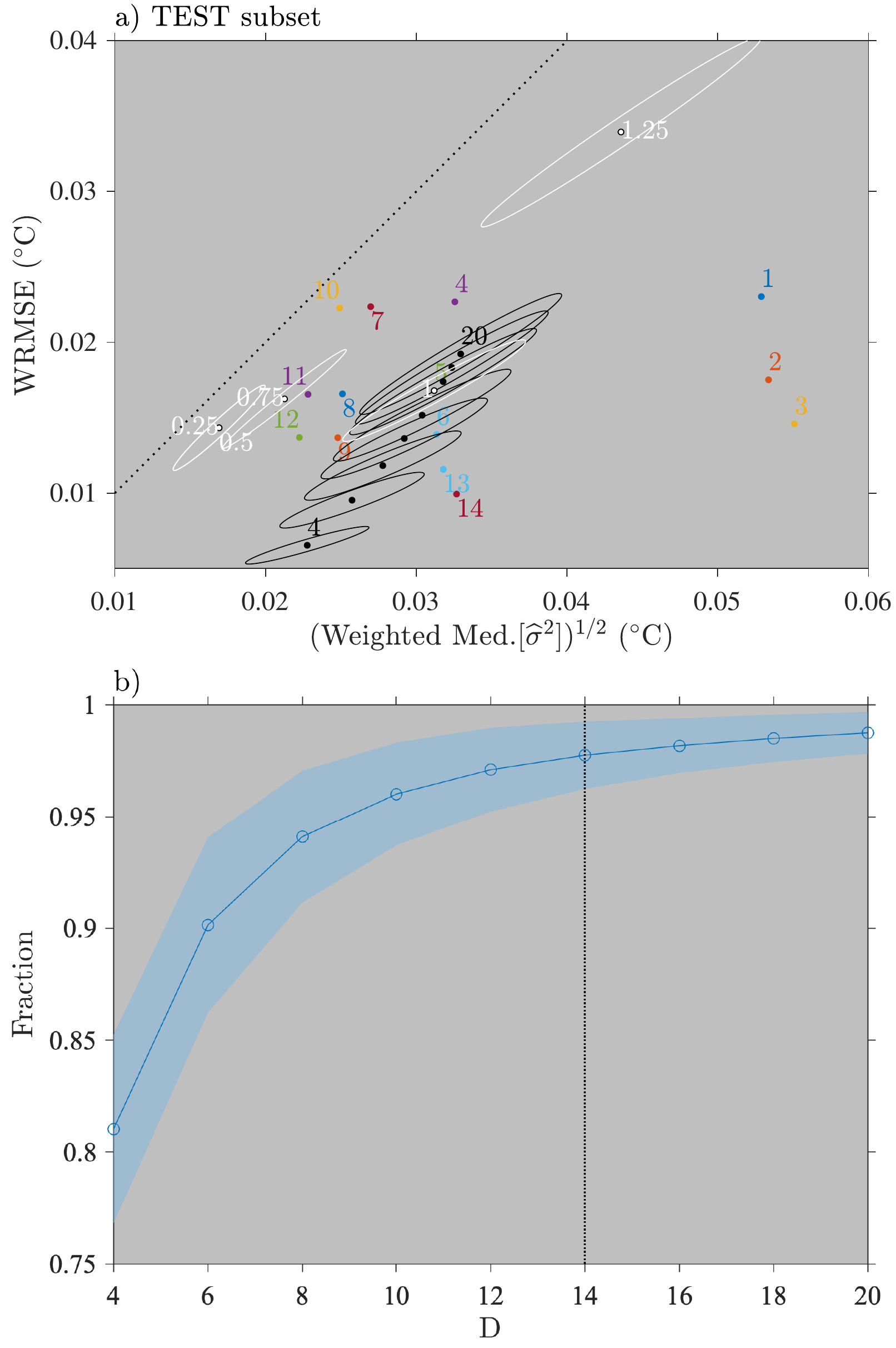}
\caption{Panel a): Summary statistics for the 14 models for the test set of drifters as in Figure~\ref{fig:modelperfs} (b). Here are also shown the results of varying the $D$ factor in eq.~(\ref{eq:delta}) from 4 to 20 in increments of 2 for model 5 (black dots and ellipses) and the results of varying the bandwidth parameter $h_k$ from 0.25 days to 1.25 days in increments of 0.25 (white dots and ellipses). The white ellipse around the black dot for model 5 corresponds to $h_k=1$ and $D=14$ as in Figure~\ref{fig:modelperfs}. Panel b): Ensemble averages of the fraction of data points not labeled as outliers as a function of factor $D$. The shading indicates plus or minus one standard deviation around the ensemble averages. The vertical dotted line indicates $D=14$ ultimately chosen here.}
\label{fig:modelperfs_test_2}
\end{figure}

% 11
\begin{figure}[ht]
\centering
\includegraphics[width=\textwidth]{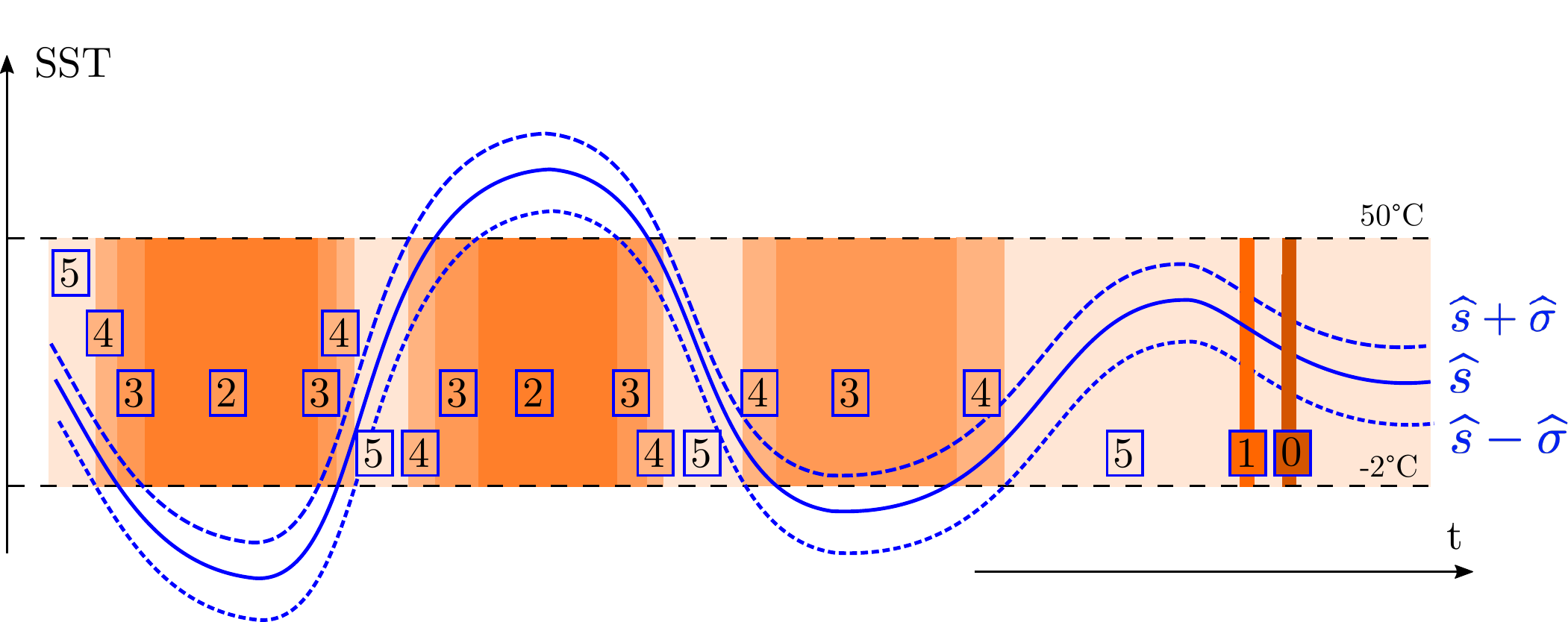}
\caption{Illustration of the quality flag determination scheme for time series of non-diurnal and total SST estimates. Flag 0 indicates both missing SST estimates ($\widehat{s}$) and uncertainty estimates ($\widehat{\sigma}$). Flag 1 indicates a missing or failed uncertainty estimate only. Flags 2, 3, 4, and 5 are based on the values of $\widehat{s}$ and $\widehat{s} \pm \widehat{\sigma}$ with respect to the range of acceptable temperature values ($[-2, 50]\,{}^\circ$C).}
\label{fig:quality_diagram}
\end{figure}

% 12
\begin{figure}[ht]
\centering
\includegraphics[width=\textwidth]{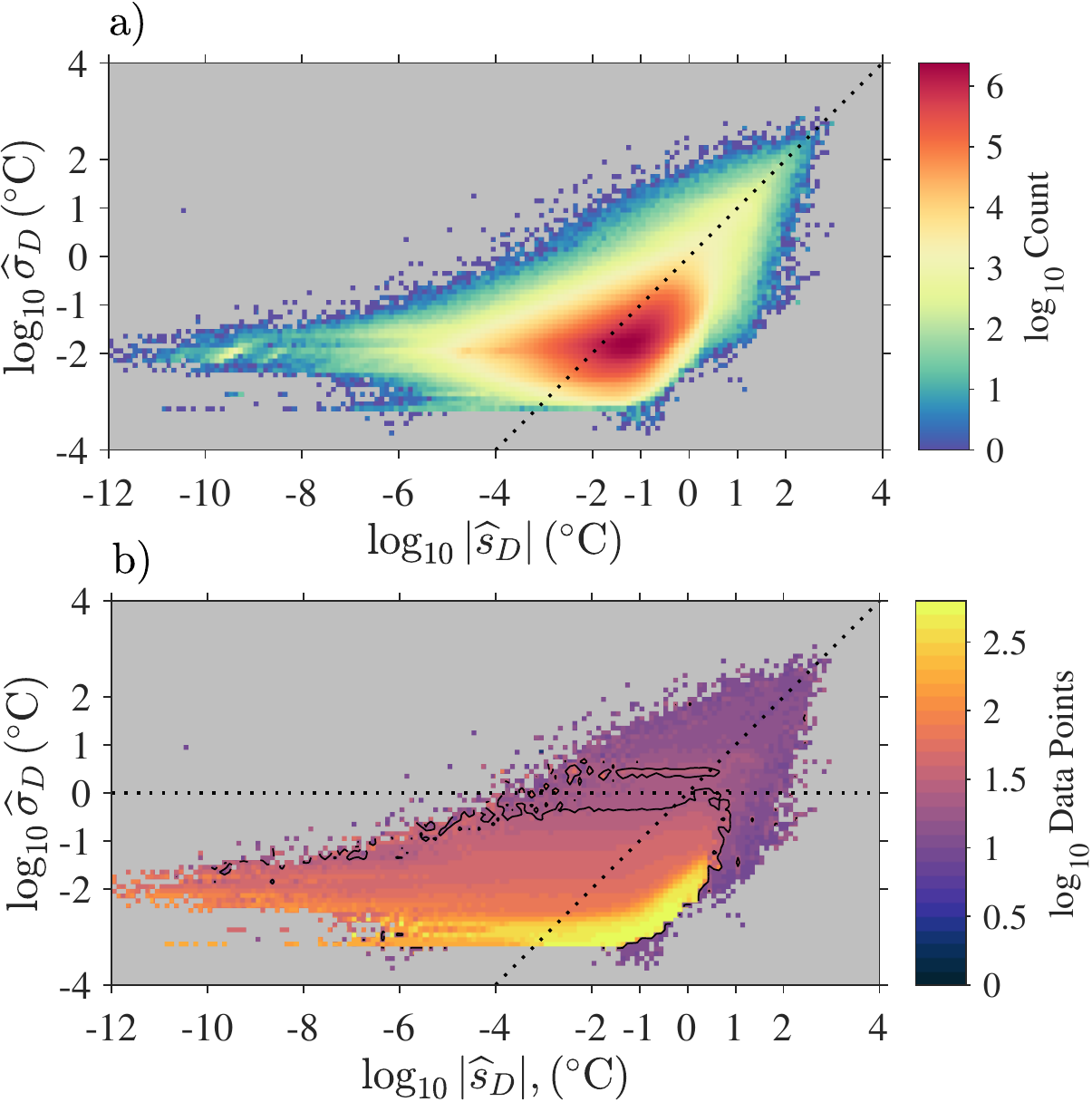}
\caption{a) Two-dimensional histogram of absolute diurnal SST estimates ($\widehat{s}_D$) and standard error estimates for diurnal estimates ($\widehat{\sigma}_D$) for {Level-3} data. The dotted line corresponds to the slope 1 line ($\widehat{\sigma}_D = \widehat{s}_D$). b) Average number of {Level-1} data points used for estimating SST mapped onto the two-dimensional distribution shown in the top panel. The black contour corresponds to 24 data points on average.}
\label{fig:sst2_esst2}
\end{figure}

% 13
\begin{figure}[ht]
\centering
\includegraphics[height=0.9\textheight]{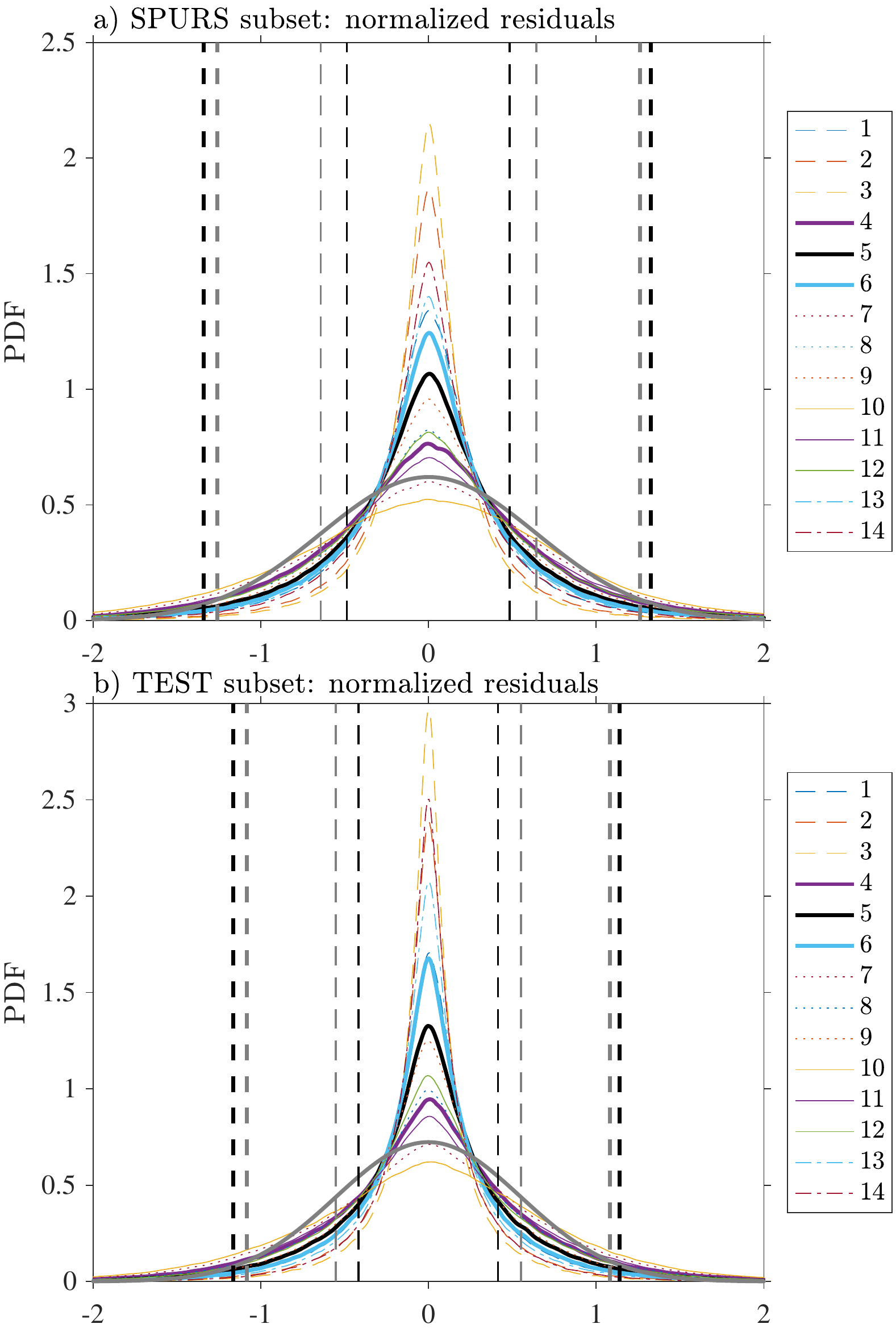}
\caption{Probability density function (PDF) estimates of normalized residuals following the fitting of the 14 models for (a) the subset of 80 drifters from the SPURS experiment and (b) the subset of 98 ``test'' drifters selected from the global database. The PDFs are estimated using an Epanechnikov kernel \cite{fan1996local} at 0.01 resolution using only residuals with non-zero final robust weights. In each panel, the thin dashed black lines indicate the 16-th and 84-th percentiles of the distribution of residuals for model 5 whereas the thick dashed black lines indicate the 2.5-th and 97.5-th percentiles. The gray curve in each panel corresponds to the fit to a normal distribution for the residuals for model 5 and the gray dashed vertical lines indicate the mean plus or minus 1 (thin line) and 1.96 (thick line) standard deviation and therefore correspond to the 2.5-, 16-, 84-, and 97.5-th percentiles of that fitted normal distribution.}
\label{fig:residuals_spurs}
\end{figure}

% 14
\begin{figure}[ht]
\centering
\includegraphics[width=\textwidth]{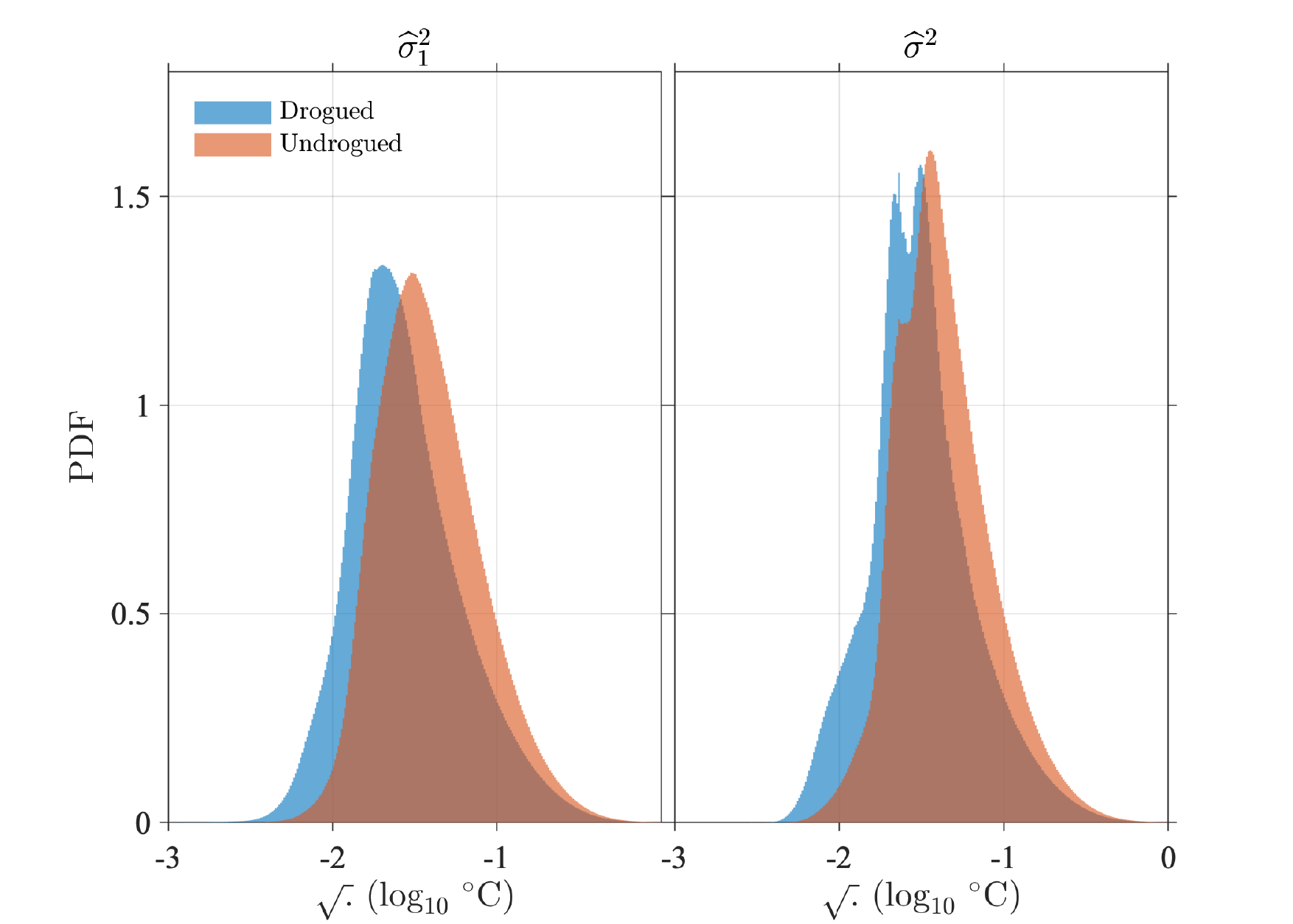}
\caption{Distribution of error variance estimates from residuals [left, $\widehat{\sigma}_1^2$, eq. (\ref{eq:wrss1})] and total error variance estimates incorporating the resolution error [right, $\widehat{\sigma}^2$, eq. (\ref{eq:wrss3})] for {Level-3} data. The histograms of the decimal logarithm of the square root of the estimates are displayed. Mode values at the peak of the distributions and 50-th percentile values are listed in Table \ref{tab:errors}.}
\label{fig:errorvar}
\end{figure}

% 15
\begin{figure}[ht]
\centering
\includegraphics[height=0.95\textheight]{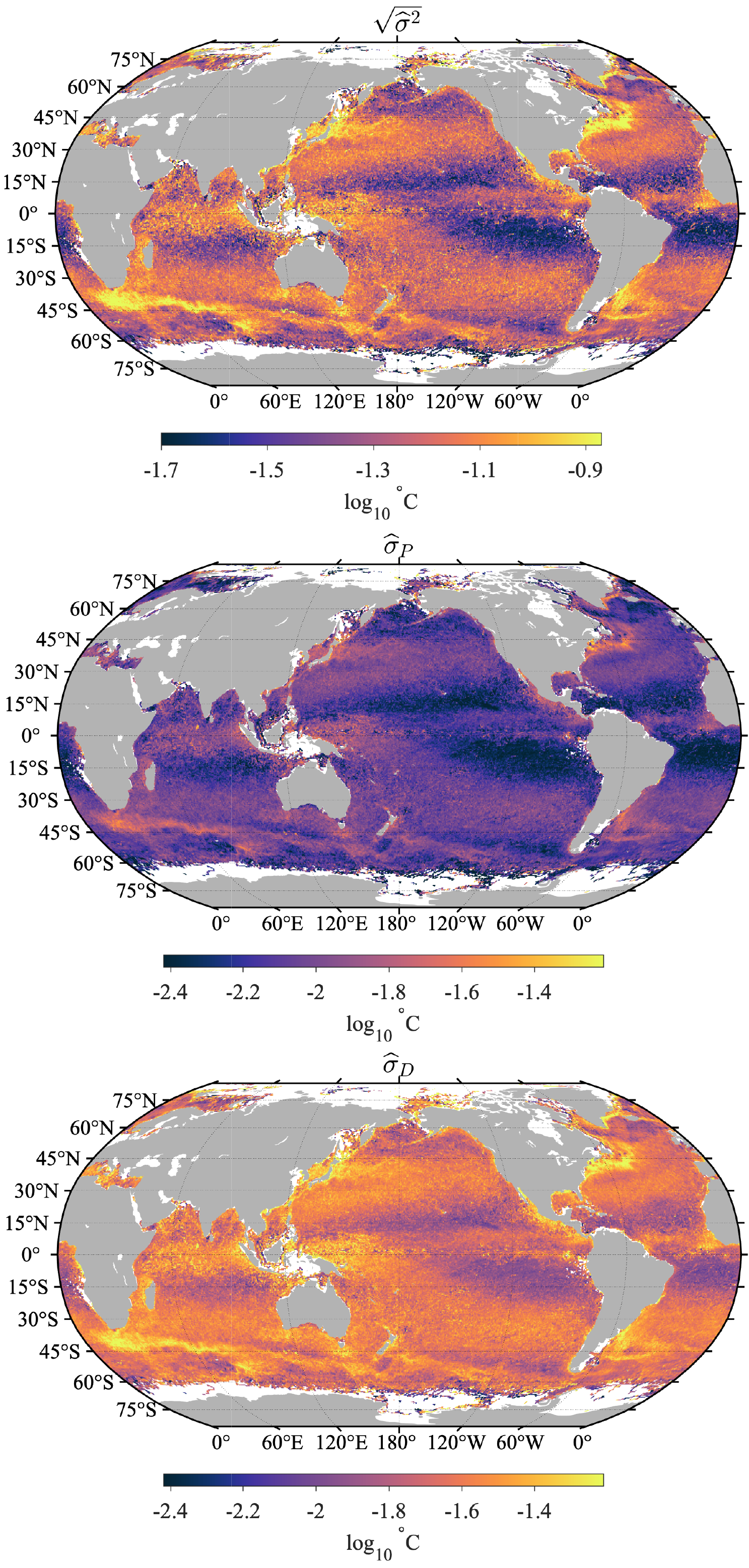}
\caption{Top: Square root of total error variance estimates [$(\widehat{\sigma^2})^{1/2}$] averaged in half-degree spatial bins. Middle: Non-diurnal SST uncertainty estimates ($\widehat{\sigma}_P$) averaged in half-degree spatial bins. Bottom: total SST uncertainty estimates ($\widehat{\sigma}_m$) averaged in half-degree spatial bins. The maps are obtained with {Level-3} data with quality flags 5 for all estimates. In all three panels the units are decimal logarithm of degrees Celsius.}
\label{fig:error_maps}
\end{figure}

% 16
\begin{figure}[ht]
\centering
\includegraphics[width=\textwidth]{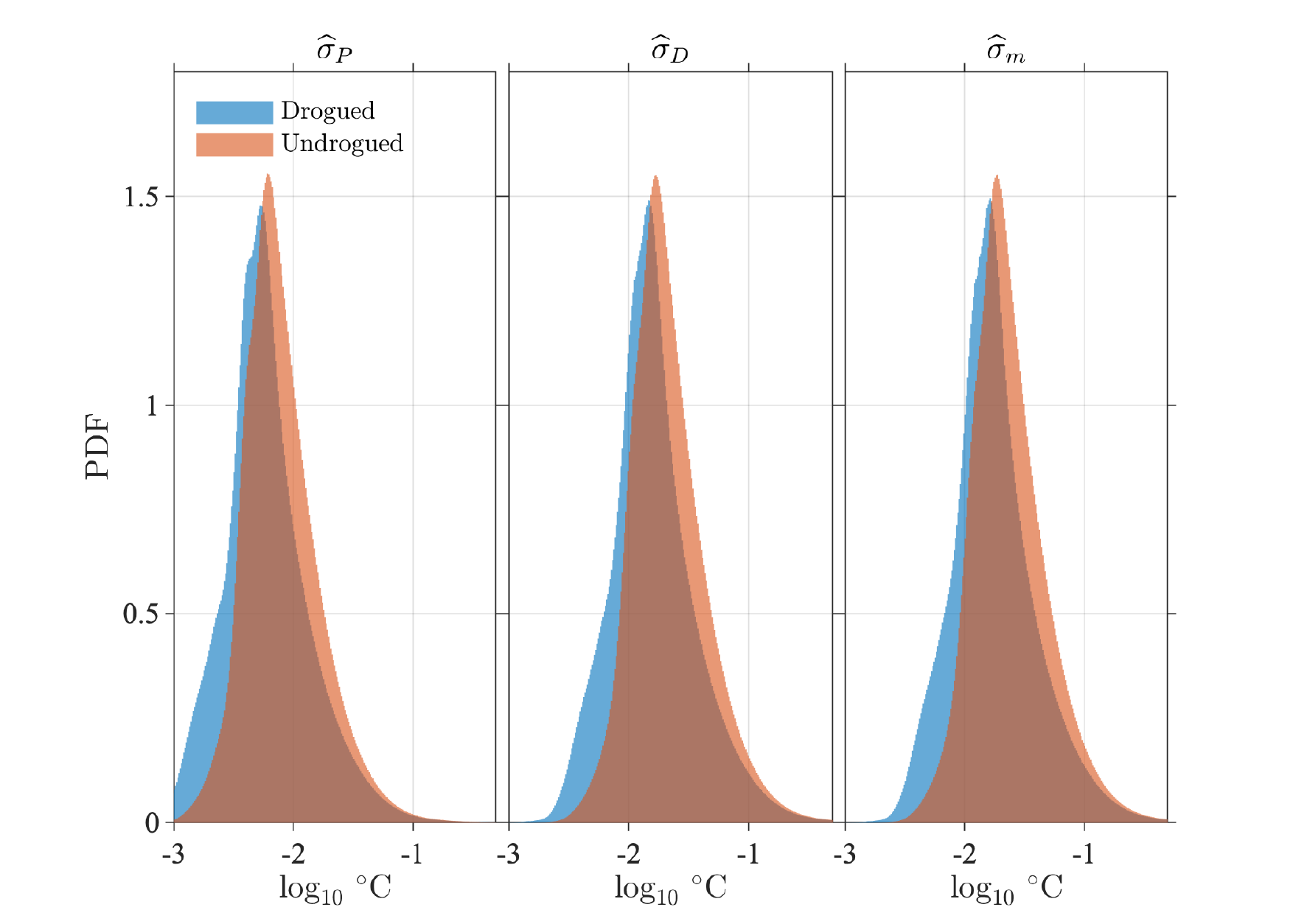}
\caption{Probability density function (PDF) estimates of standard error estimates for the non-diurnal ($\widehat{\sigma}_P$), diurnal ($\widehat{\sigma}_D$), and total ($\widehat{\sigma}_m$), SST estimates, separated between data from drogued and undrogued drifters. The normalized histograms of the decimal logarithm of the estimates are displayed. Mode values at the peak of the distributions and 50-th percentile values are listed in Table \ref{tab:uncertainties}.}
\label{fig:esst}
\end{figure}

% 17
\begin{figure}[ht]
\centering
\includegraphics[height=0.95\textheight]{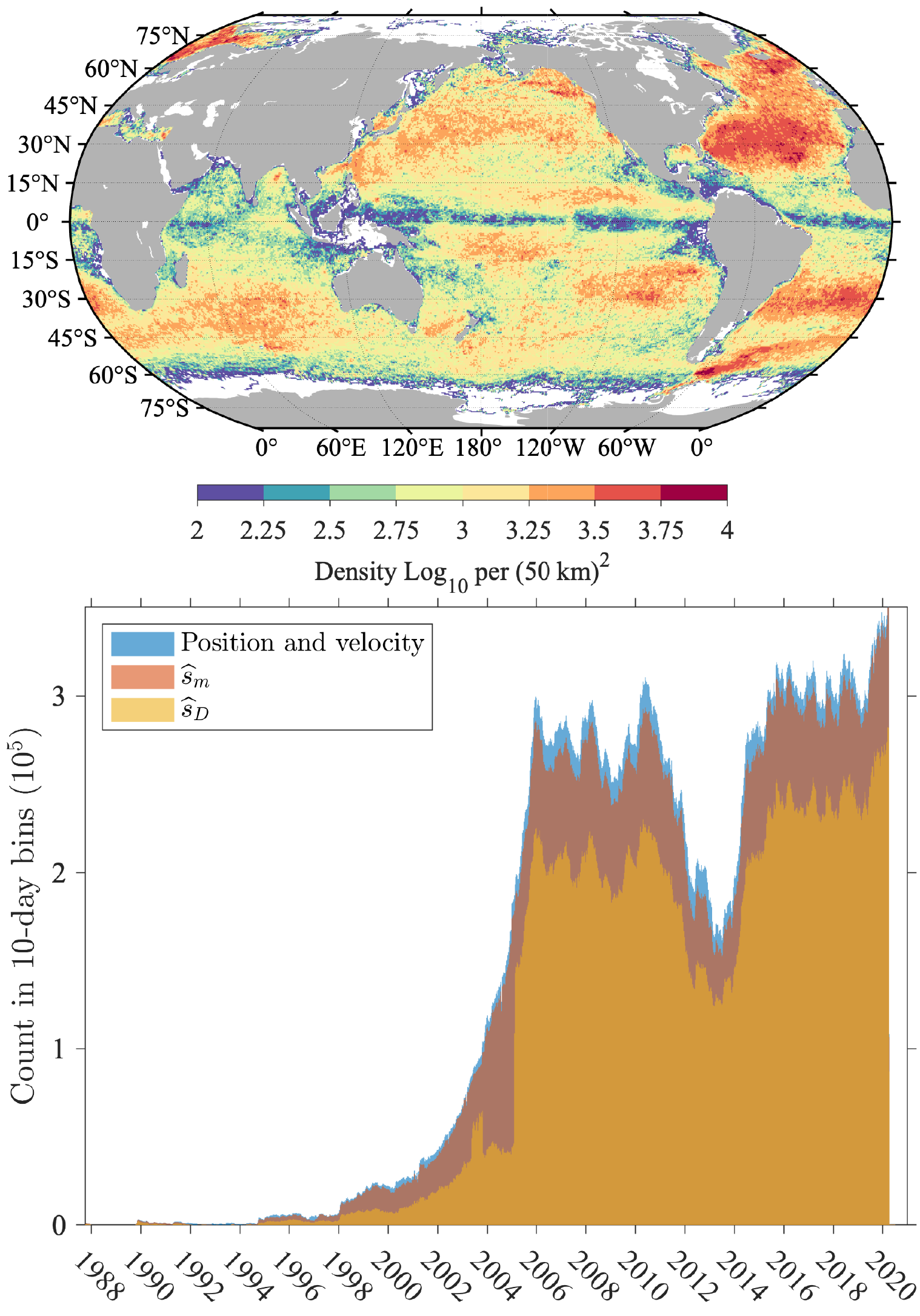}
\caption{Top panel: Spatial distribution of {Level-3} total SST estimates expressed as a density per (50 km)$^2$ in half-degree spatial bins. Only quality flag 5 data are counted. Bottom panel: {Level-3} total ($\widehat{s}_m$) and diurnal ($\widehat{s}_D$) SST estimates temporal distribution in 10-day bins from 03-Oct-1987 13:00:00 to 30-Jun-2020 23:00:00. The temporal distribution of the matching position and velocity hourly dataset \cite{elipot2016global} release 2 is also displayed. The temporal distribution of non-diurnal SST estimates is not displayed as it would be indistinguishable from the distribution of the total SST estimates.}
\label{fig:density_histogram}
\end{figure}

% 18
\begin{figure}[ht]
\centering
\includegraphics[height=0.95\textheight]{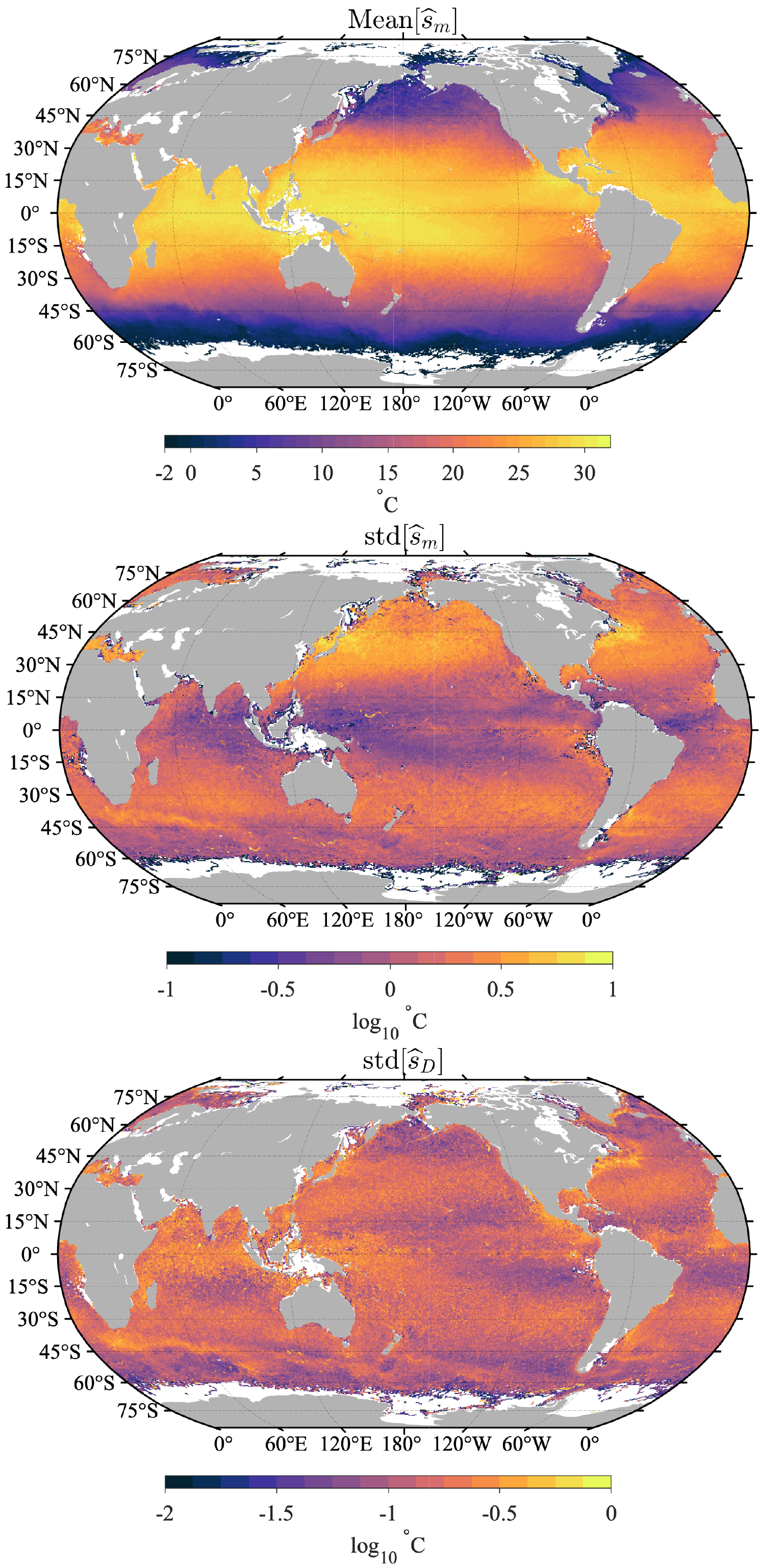}
\caption{Top: {Level-3} total SST estimates averaged in half-degree spatial bins. Middle: {Level-3} total SST estimates standard deviation in half-degree spatial bins. Bottom: {Level-3} diurnal SST estimates standard deviation in half-degree spatial bins. Only quality flag 5 estimates for each respective variable is used to produce these maps.}
\label{fig:mean_std}
\end{figure}

% 19
\begin{figure}[ht]
\centering
\includegraphics[width=\textwidth]{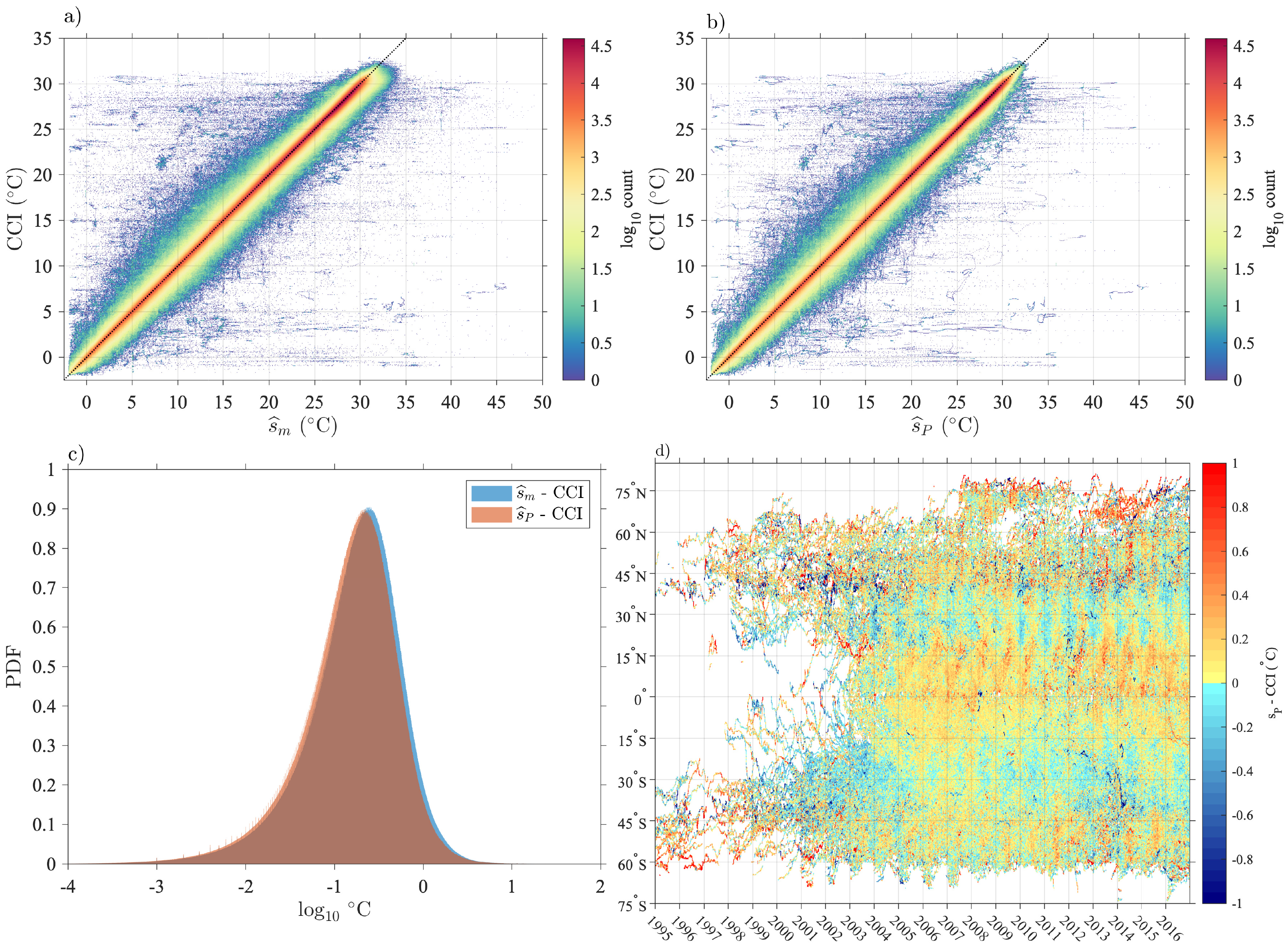}
\caption{{Comparison between the drifter hourly SST dataset and the ESA SST CCI Analysis v2.1 product}\cite{good2019esa,merchant2019satellite}{. a) Two-dimensional histogram in 0.05$^\circ$C bins of drifter total SST estimates ($\widehat{s}_m$) versus the interpolated CCI values. b) Same as in a) but for the drifter non-diurnal SST estimates ($\widehat{s}_P$). c) Normalized histograms of absolute differences between drifter estimates and CCI values. d) Average differences between non-diurnal SST estimates ($\widehat{s}_P$) and CCI values as a function of time and latitude.}}
\label{fig:sst_vs_cci}
\end{figure}

% 20
\begin{figure}[ht]
\centering
\includegraphics[width=\textwidth]{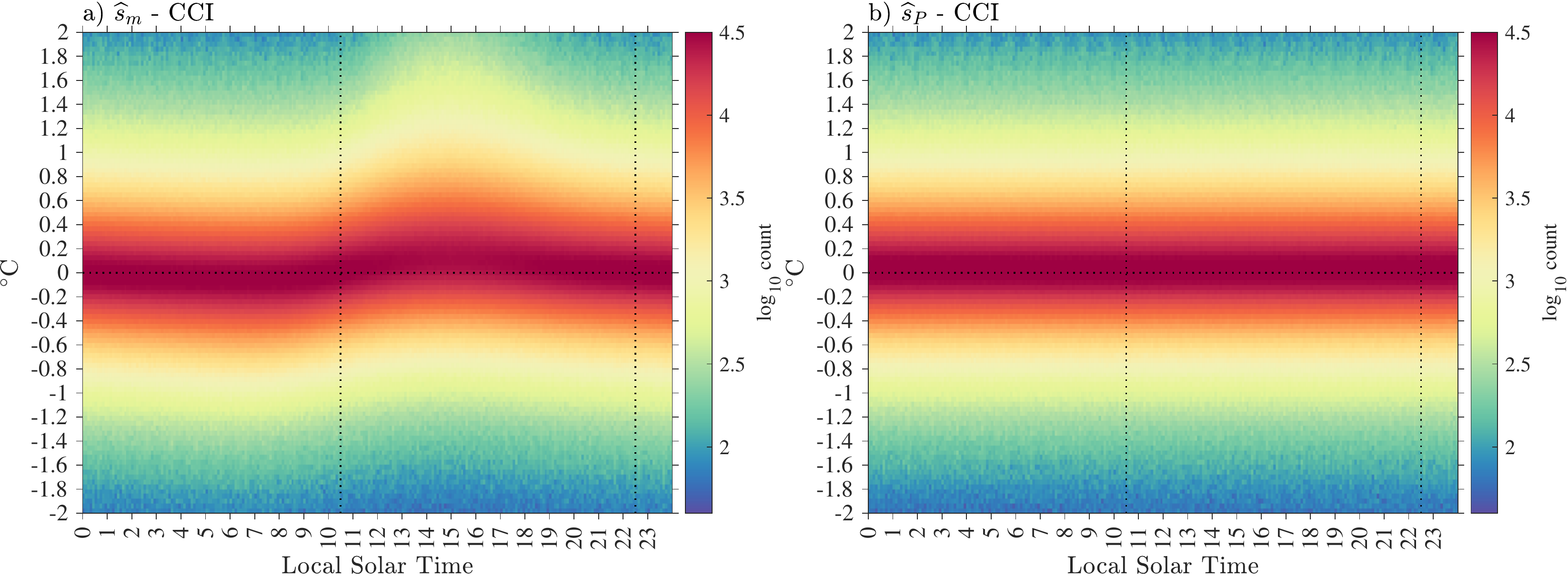}
\caption{{a) Two-dimensional histogram of drifter total SST estimates minus CCI values ($\widehat{s}_m$-CCI) versus local mean solar time. b) Same as in a) but for drifter non-diurnal SST estimates ($\widehat{s}_P$-CCI). In both panels the vertical dotted lines indicate 10:30 and 22:30 local mean solar times.}}
\label{fig:sst_minus_cci_lst}
\end{figure}

\clearpage

\begin{sidewaystable}
  \centering
  \begin{tabular}{lccccc}
\multicolumn{2}{c}{}    & \multicolumn{2}{c}{Argos drifters} & \multicolumn{2}{c}{Iridium drifters} \\
\cline{3-6}
\multicolumn{2}{c}{} & Location $(X,Y)$ \& velocity $(U,V)$ & Temperature ($S$) & Location $(X,Y)$ \& velocity $(U,V)$ & Temperature ($S$) \\
\hline
\multicolumn{2}{l}{{Level-1}} & \shortstack[c]{\\$X,Y$} & $S$ & $X,Y$ & $S$ \\
\hline
\multicolumn{2}{l}{{Level-2}} & - & \shortstack[c]{$S^*$\\ \emph{This paper}} & \shortstack[c]{\\$X^*,Y^*,U^*,V^*$\\ \emph{Elipot et al. (2016)}} & \shortstack[c]{$S^*$\\ \emph{This paper}} \\
\hline
{{Level-3}} & 6-hourly & \shortstack[c]{\\$\mathbf{X^*,Y^*,U^*,V^*}$\\ \emph{Hansen and Poulain (1996)}} & \shortstack[c]{$\mathbf{S^*}$\\ \emph{Hansen and Poulain (1996)}} & \shortstack[c]{$\mathbf{X^*,Y^*,U^*,V^*}$\\ \emph{Hansen and Poulain (1996)}} & \shortstack[c]{$\mathbf{S^*}$\\ \emph{Hansen and Poulain (1996)}} \\
\cline{2-6} 
& hourly & \shortstack[c]{\\$\mathbf{X^*,Y^*,U^*,V^*}$\\ \emph{Elipot et al. (2016)}} & \shortstack[c]{$\mathbf{S^*}$\\ \emph{This paper}} & \shortstack[c]{$\mathbf{X^*,Y^*,U^*,V^*}$\\ \emph{Elipot et al. (2016)}} & \shortstack[c]{$\mathbf{S^*}$\\ \emph{This paper}} \\
\cline{2-6}
  \end{tabular}
  \caption{Table of location, velocity, and temperature data products and availability from the Global Drifter Program (GDP) as described in Hansen and Poulain (1996), Elipot et al. (2016), and this paper which defines three levels of data processing. Variables without asterisks are observations and variables with asterisks are estimates. Only {Level-3} data products are readily available from the GDP.}
  \label{tab:datamatrix}
\end{sidewaystable}

\begin{table}
\centering
\begin{tabular}{ccccc}
\hline\hline
Polynomial order (P)  & 0 & 1 & 2  & 3\\
\hline
Diurnal harmonics (N) & & \\
 2 & $\mathbf{1}$ (5) &   $\mathbf{4}$ (6) & $\mathbf{7}$ (7) & $\mathbf{10}$ (8)\\
 3 & $\mathbf{2}$ (7) &   $\mathbf{5}$ (8) & $\mathbf{8}$ (9) & $\mathbf{11}$ (10)\\
 4 & $\mathbf{3}$ (9) &   $\mathbf{6}$ (10) & $\mathbf{9}$ (11) & $\mathbf{12}$ (12)\\
 5 &  &   $\mathbf{13}$ (12) &  & \\
 6 & &   $\mathbf{14}$ (14) &  & \\
\hline
\end{tabular}
\caption{Table of temporal models considered for SST as a function of polynomial order and number of diurnal harmonics. Boldface numbers indicate the model identification numbers discussed in the text. Number in parentheses indicate the number of parameters of each model ($2N+P+1$). The model ultimately chosen is number 5.}
\label{tab:models}
\end{table}

\begin{table}
\centering
\begin{tabular}{ccccccc}
\hline
Data Flag  & 0 & 1 & 2  & 3 & 4 & 5\\
 \hline\hline
 Total ($\widehat{s}_m$) &  6,909,407 & 302,743 & 80,542 & 10,458 & 14,413 & 158,436,770 \\
                                      & 4.17\% & 0.18\% & 0.05\% & 0.01\% & 0.01\% & 95.59\% \\
\hline
Non-diurnal ($\widehat{s}_P$) &  6,909,407 & 302,743 & 77,844 & 4,425 & 6,823 & 158,453,091 \\
                                      & 4.17\% & 0.18\% & 0.05\% & <0.01\% & <0.01\% & 95.60\% \\
\hline
Diurnal ($\widehat{s}_D$) &  6,909,407 & 302,743 & 34,391,616 & 51,181 & 3,787,063 & 120,312,323 \\
                                      & 4.17\% & 0.18\% & 20.75\% & 0.03\% & 2.28\% & 75.58\% \\
\hline
\end{tabular}
\caption{Inventory of quality flags for {Level-3} estimates. The target number of data points is 165,754,333 from 17,324 time series for the Global Drifter program hourly dataset version 2.0.}
\label{tab:quality}
\end{table}

\begin{table}
\centering
\begin{tabular}{llccc}
\hline
\hline
$\sqrt{\widehat{\sigma}^2_1}$ & & All & Drogued & Undrogued \\ 
 \hline
Mode   & {Level-2} & 0.026 & 0.020 & 0.030 \\
             & \textbf{{Level-3}} & \textbf{0.026} & \textbf{0.020} & \textbf{0.030} \\
\hline
50-th percentile & {Level-2} & 0.031 & 0.025 & 0.036 \\
                & \textbf{{Level-3}} & \textbf{0.031} & \textbf{0.025} & \textbf{0.036} \\
% \hline
% 0.5-th percentile & Level-1 & 0.005 & 0.004 & 0.007 \\
%                 & Level-2 & 0.006 & 0.005 & 0.008 \\
% \hline
% 99.5-th percentile & Level-1 & 0.340 & 0.329 & 0.347 \\
%                 & Level-2 & 0.332 & 0.312 & 0.343 \\
\hline
\hline
$\sqrt{\widehat{\sigma}^2}$ & & All & Drogued & Undrogued \\ 
 \hline
Mode   & {Level-2} & 0.033 & 0.031 & 0.036 \\
             & \textbf{{Level-2}} & \textbf{0.033} & \textbf{0.031} & \textbf{0.036} \\
\hline
50-th percentile & {Level-2} & 0.035 & 0.030 & 0.040 \\
                & \textbf{{Level-3}} & \textbf{0.036} & \textbf{0.030} & \textbf{0.040} \\
% \hline
% 0.5-th percentile & Level-1 & 0.007 & 0.006 & 0.009 \\
%                 & Level-2 & 0.007 & 0.006 & 0.009 \\
% \hline
% 99.5-th percentile & Level-1 & 0.340 & 0.329 & 0.347 \\
%                 & Level-2 & 0.332 & 0.312 & 0.343 \\
\hline
\end{tabular}
\caption{Statistics of error variance estimates from residuals [${\widehat{\sigma}^2_1}$, eq.~(\ref{eq:wrss1})] and final error variance estimates incorporating the resolution error [${\widehat{\sigma}^2}$, eq.~(\ref{eq:wrss3})]. The square root values are displayed, rounded to the nearest 0.001. Units are degrees Celsius. Bold values highlights the values discussed in the text: the 50-th percentile (median) and mode values of error standard deviation for {Level-3} data.}
\label{tab:errors}
\end{table}

\begin{table}
\centering
\begin{tabular}{llccc}
\hline\hline
$\widehat{\sigma}_P$  & & All & Drogued & Undrogued \\ 
 \hline
Mode   & {Level-2} & 0.006 & 0.005 & 0.006 \\
             & \textbf{{Level-3}} & \textbf{0.006} & \textbf{0.005} & \textbf{0.006} \\
\hline
50-th percentile & {Level-2} & 0.006 & 0.006  & 0.007  \\
                & \textbf{{Level-3}} & \textbf{0.007} & \textbf{0.006} & \textbf{0.007} \\
\hline\hline
$\widehat{\sigma}_D$ & & All & Drogued & Undrogued \\ 
 \hline
Mode   & {Level-2} &0.016 &0.015 & 0.016\\
             & \textbf{{Level-3}} & \textbf{0.016} & \textbf{0.015} & \textbf{0.017} \\
\hline
50-th percentile & {Level-2} & 0.017& 0.015&0.019 \\
                & \textbf{{Level-3}} & \textbf{0.018} & \textbf{0.015} & \textbf{0.020} \\
\hline\hline
$\widehat{\sigma}_m$  & & All & Drogued & Undrogued \\ 
 \hline
Mode   & {Level-2} & 0.017 & 0.016 & 0.018 \\
             & \textbf{{Level-3}} & \textbf{0.018} & \textbf{0.016} & \textbf{0.019} \\
\hline
50-th percentile & {Level-2} & 0.018& 0.016& 0.020 \\
                & \textbf{{Level-3}} & \textbf{0.019} & \textbf{0.016} & \textbf{0.022} \\
\end{tabular}
\caption{Statistics of SST standard uncertainty estimates. Units are degrees Celsius. Values are rounded to the nearest 0.001.}
\label{tab:uncertainties}
\end{table}

\begin{sidewaystable}
\centering
\begin{tabular}{ll}
\hline \hline
Description of the content of variables in the NetCDF files & Names of variables \\
\hline \hline
\shortstack[l]{Time of estimate} & time \\
\hline
\shortstack[l]{Estimated latitude \\ from drifting locations} & latitude \\
\hline
\shortstack[l]{Estimated longitude \\ from drifting locations} & longitude \\
\hline
\shortstack[l]{Estimated northward velocity \\ from drifting locations} & surface northward sea water velocity \\
\hline
\shortstack[l]{Estimated eastward velocity \\ from drifting locations} & surface eastward sea water velocity \\
\hline
\shortstack[l]{Estimated 95\% confidence interval in latitude \\ from drifting locations} & 95\% confidence interval in latitude \\
\hline
\shortstack[l]{Estimated 95\% confidence interval in longitude \\ from drifting locations} & 95\% confidence interval in longitude \\
\hline
\shortstack[l]{Estimated 95\% confidence interval in northward velocity \\ from drifting locations} & 95\% confidence interval in northward velocity \\
\hline
\shortstack[l]{Estimated 95\% confidence interval in eastward velocity \\ from drifting locations} & 95\% confidence interval in eastward velocity \\
\hline
\shortstack[l]{Estimated near-surface sea water temperature \\ from drifting buoy measurements} & fitted sea water temperature \\
\hline
\shortstack[l]{Estimated near-surface non-diurnal sea water temperature \\ from drifting buoy measurements} & fitted non-diurnal sea water temperature \\
\hline
\shortstack[l]{Estimated near-surface diurnal sea water temperature anomaly \\ from drifting buoy measurements} & fitted diurnal sea water temperature anomaly \\
\hline
\shortstack[l]{Estimated one standard error of near-surface sea water temperature \\ estimate from drifting buoy measurements} & \shortstack[l]{standard uncertainty of fitted \\ sea water temperature}\\
\hline
\shortstack[l]{Estimated one standard error of near-surface non-diurnal sea water temperature \\ estimate from drifting buoy measurements} & \shortstack[l]{standard uncertainty of fitted non-diurnal \\ sea water temperature} \\
\hline
\shortstack[l]{Estimated one standard error of near-surface diurnal sea water temperature anomaly \\ estimate from drifting buoy measurements} & \shortstack[l]{standard uncertainty of fitted diurnal \\ sea water temperature anomaly} \\
\hline
 Quality flag for fitted sea water temperature & fitted sea water temperature quality flag\\
\hline
 Quality flag for fitted non-diurnal sea water temperature & fitted non-diurnal sea water temperature quality flag\\
\hline
 Quality flag for fitted diurnal sea water temperature anomaly & fitted diurnal sea water temperature anomaly quality flag\\
\hline
\end{tabular}
\caption{Data record information for the {Level-3} data product and details of variables in drifter NetCDF files.}
\label{tab:detail}
\end{sidewaystable}

\end{document}